\documentstyle[12pt,a4wide,epsf,rotate]{article}

\newcommand{\beq}{\begin{eqnarray}}
\newcommand{\eeq}{\end{eqnarray}}
\newcommand{\lt}{\left}
\newcommand{\r}{\right}
\newcommand{\mc}{m_c^2}
\newcommand{\mt}{m_t^2}
\newcommand{\ms}{m_s^2}
\newcommand{\md}{m_d^2}
\newcommand{\mw}{M_W^2}
\newcommand{\oll}{Q_{LL}}
\newcommand{\ov}{\overline}
\newcommand{\nn}{\nonumber \\}
\newcommand{\no}{\nonumber }
\newcommand{\tw}{\hat{\tilde{1}}}
\newcommand{\1}{\hat{1}}
\newcommand{\T}{{\, \rm \bf T \,}}
\newcommand{\gev}{\, {\rm GeV}}
\newcommand{\laMSb}{\Lambda_{\overline {\rm MS}}}
\newcommand{\gf}{\, {\rm G_F}}
\newcommand{\fig}[1]{fig.~\ref{#1}}

\begin{document}

%
%
\title{\sf
	{\large
		\rightline{TUM-T31-49/93}
		\rightline{hep-ph/9310311}
		\rightline{October 1993}
	}
	\vspace{1cm}
	{\LARGE
	Enhancement of the $K_L$ - $K_S$ Mass Difference
	by Short Distance QCD Corrections Beyond
	Leading Logarithms
	\thanks{Supported by the German Bundesministerium f\"{u}r Forschung
		und Technologie under the contract 06~TM~732.}
	}
}
\author{
	Stefan Herrlich
	\thanks{e-mail: {\tt herrl@feynman.t30.physik.tu-muenchen.de} }
	\hspace{0.2cm} and \hspace{0.2cm}
	Ulrich Nierste
	\thanks{e-mail: {\tt nierste@feynman.t30.physik.tu-muenchen.de} }
	\thanks{supported in part by Studienstiftung des Deutschen Volkes}
	\\[3mm]
	{\small \sl Physik-Department T31,
		Technische Universit\"{a}t M\"{u}nchen,
		D-85747 Garching, FRG}
}
\date{}

\maketitle
\vspace{2cm}

%
%
\centerline{\large \bf Abstract}
\vspace{1cm}
We calculate the next-to-leading order short distance QCD corrections to the
coefficient $\eta_1$ of the effective $\Delta S = 2$ hamiltonian in the
standard model.
This part dominates the short distance contribution $(\Delta m_K)^{\rm SD}$
to the $K_L$ -- $K_S$ mass difference.
The next-to-leading order result enhances $\eta_1$ and
$(\Delta m_K)^{\rm SD}$ by 20\% compared to the leading order estimate.
Taking
$0.200 \gev \le \laMSb \le 0.350 \gev$
and
$1.35 \gev \le m_c(m_c) \le 1.45 \gev$
we obtain
$0.922 \le \eta_1^{\rm NLO} \le 1.419$
compared to
$0.834 \le \eta_1^{\rm LO} \le 1.138$.
For $B_K = 0.7$ this corresponds to 48 -- 75 \% of the experimentally
observed mass difference.
The inclusion of next-to-leading order corrections to $\eta_1$ reduces
considerably the theoretical uncertainty related to the choice of
renormalization scales.

\thispagestyle{empty}
\newpage
\setcounter{page}{1}
%
%

%
%
\section{Introduction}

Over the last three decades the study of the K--meson system has contributed
enormously to our insight into the fundamental principles of nature,
just to mention the discovery of CP--violation  
by Christensen et al.\ \cite{c}
and the conjecture of the existence of the charm quark by Glashow,
Iliopoulos and Maiani \cite{gim}. 
The latter was derived from the suppression of flavour--changing neutral
currents, which is responsible for
the smallness of $K^{0}-\overline{K^0}$--mixing.
In the following years this
$\Delta S=2$ weak interaction process has been used to investigate the
indirect CP--violation apparent in  the $K_{L}-K_{S}$ mass difference
and to constrain the parameters of the Standard Model.
The first highlight  was clearly the
prediction of  the mass of the charm quark
prior to its discovery by Gaillard and Lee \cite{gl}.

The theoretical determination of the effective low energy
$\Delta S=2$--hamiltonian, however, is not satisfactory yet, because the
indispensable incorporation of strong interaction effects is difficult due
to the poorly known infrared structure of quantum chromodynamics.
The theorist's tool to tackle the problem is the use of Wilson's operator
product expansion factorizing the Feynman amplitudes into a long distance
part, which has to be calculated by non--perturbative methods, and a short
distance Wilson coefficient,
to which perturbative methods are applicable. The largely
separated mass scales in the problem generate large logarithms, whose
resummation is mandatory. This is achieved by renormalization group methods
applied to a cascade of effective field theories, in which the heavy
degrees of freedom are successively removed.

Pioneering work in the short distance calculation of the
$\Delta S=2$--hamiltonian
was done by
Vysotskij \cite{v} and by
Gilman and Wise \cite{gw}, who have used the leading
logarithmic (LL)
approximation. This method leaves certain questions, which will
be summarized below, unanswered,
and therefore many attempts have been made to include subleading effects.

To be specific, consider the hamiltonian:
\begin{eqnarray}
\lefteqn{H^{\Delta S=2}(\mu)=} \nn
&& \hspace{-5ex} \frac{G_{F}^2}{16 \pi^2} \mw \!\! \lt[
                  \lambda_c^2 \eta_1
                  S(\frac{\mc }{\mw }) \! + \!
                  \lambda_t^2 \eta_2
                  S(\frac{\mt }{\mw }) \! + \!
                2 \lambda_c \lambda_t \eta_3
                  S(\frac{\mc }{\mw },
                  \frac{\mt }{\mw })
                   \r]  \! \!
 b(\mu) \oll(\mu), \; \;\label{s2}
\end{eqnarray}
where $G_{F}$
is the Fermi constant, $\lambda_j=V_{jd} V_{js}^{*}$ denotes the CKM--factors,
and
$\oll$ is  the local
four quark operator (see \fig{loc} on p.\ \pageref{loc})
\begin{eqnarray}
\oll &=&
( \ov{s}_j \gamma_\mu (1-\gamma_5) d_j)
(\ov{s}_k \gamma^\mu (1-\gamma_5) d_k) \; = \; (\ov{s} d)_{V-A}
(\ov{s} d)_{V-A}  \label{oll}
\end{eqnarray}
with j and k being colour indices.
%

\begin{figure}[ht]
\centerline{
\rotate[r]{
\epsfysize=5cm
\epsffile{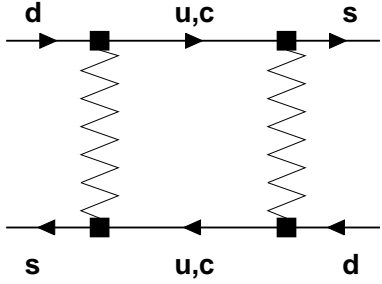}
}}
\caption[]{
	The lowest order box diagrams contributing to $H^{\Delta S=2}$.
	The zigzag lines stand for W--bosons or
	ficticious Higgs particles.}
\label{box}
\end{figure}

In eq.\ (\ref{s2}) the GIM mechanism
$\lambda_u+\lambda_c+ \lambda_t=0$ has been used to eliminate $\lambda_u$.
$S(x,y)$ and $S(x)$
denote the Inami--Lim functions \cite{il} resulting from the evaluation
of the box diagrams of \fig{box}.
They must be understood to be expanded to lowest
nonvanishing order
in $\mc/\mw$ in  (\ref{s2}).

The $\eta_i$'s contain the QCD corrections with their dependence
on the renormalization scale $\mu $
explicitly factored out in the coefficient
$b(\mu)$. In the LL approximation $b(\mu)$ reads in terms of the
three--flavour strong
interaction
constant:
$b(\mu)= \lt[ \alpha_3(\mu) \r]^{-2/9} $.
In the absence of QCD corrections
$\eta_i b(\mu) =1$. Since physical quantities must not depend on
$\mu$, $b(\mu)$  has to  cancel the corresponding dependence in the
matrix element of $\oll$, which reads in the case of
$K^{0}-\overline{K^0}$--mixing
\begin{eqnarray}
\langle   \ov{K^0} \mid \oll (\mu) \mid {K^0} \rangle &=&
\frac{8}{3} B_K (\mu ) f_K^2 m_K^2 .
\end{eqnarray}
Here $m_K$ and $f_K$ are the mass and decay constant of the $K$--meson,
and the `bag'--parameter $B_K(\mu)$ combines with $b(\mu)$ to the
renormalization group invariant
\begin{eqnarray}
B_K &=& B_K (\mu ) b( \mu ). \no
\end{eqnarray}

In the following we mention the main reasons for going beyond the
LL approximation in the calculation of $\eta_1$:
\begin{itemize}
\item The running charm quark mass $m_c$ enters
      $H^{\Delta S=2}$ at the scale $\mu_c \simeq m_c$,
      at which the effective four--quark
      theory is matched to an effective three--quark theory.
      The LL result depends considerably on $\mu_c$.
\item Similarly $\eta_1$ depends on the scale $\mu_W \simeq M_W$,
      at which the W--boson is removed from the calculation.
\item The definition of the QCD scale parameter $\Lambda$ requires a
      next--to--leading order calculation, and therefore
      one can at most qualitatively discuss the dependence on $\Lambda $
      of physical quantities calculated in the LL approximation.
\item After all the subleading terms can be  sizeable.
      The LL hamiltonian only explains about  60~\% of the observed
      $K_L-K_S$ mass difference.
\end{itemize}

Buras,  Jamin and  Weisz \cite{bjw} have improved
the result of \cite{gw} by including
next--to--leading order (NLO) effects into $\eta_2$
for $H^{\Delta S=2}$
and
$H^{\Delta B=2}$,
which,
being sensitive to the unknown top--quark mass,
is relevant for indirect CP violation in K--decays and
$B^0-\ov{B^0}$-mixing.
They have thoroughly
verified that
their Wilson coefficient depends neither on the gauge of the
gluon propagator nor on the infrared structure of the factorized
Feynman amplitude and proved  previous attempts to
incorporate NLO effects into
$H^{\Delta S=2}$,
$H^{\Delta B=2}$
to be incorrect.
Since the present work extends theirs to the coefficient $\eta_1$,
we closely stick to the notation in \cite{bjw}.

An earlier attempted  NLO calculation \cite{fky} of $\eta_1$ has already
been commented on in \cite{bjw}. The authors of \cite{fky} have incorrectly
extracted the short distance part of the transition amplitude and have thereby
found a drastic decrease in $\eta_1$ rather than the
sizeable increase found by us.

The coefficient $\eta_1$ plays the key role for the
$K_L-K_S$ mass difference, while $\eta_2$ and $\eta_3$
contribute only   less than  10\%.
Since the LL calculation has  yielded a too small  mass
difference, the correction in \cite{fky}
even worsens the LL result.

The nonperturbative parameter $B_K$ has been calculated using lattice
gauge theories \cite{gks}, $1/N$ expansion \cite{bbg} and QCD sum rules
\cite{d} leading to  $B_K=0.7 \pm 0.2$. The QCD hadron
duality approach   favours lower values $B_K=0.4 \pm 0.1$ \cite{pr}.
For a large $N$ calculation of both short and long
distance contributions to the $K_L-K_S$--mass difference see \cite{gb}.

Our paper is organized as follows: In section 2 we will set up our notations
and present the general
outline of  the  calculations.
Section 3 briefly
summarizes the LL calculation of \cite{gw} changing the assumption
$m_t \ll M_W$ into the present day relation
$m_t \parbox[c]{1.8ex}{$\stackrel{>}{\sim}$} M_W$.
Section 4 is devoted to our NLO calculation followed by section 5 containing
the numerical analysis of our results. We close  the paper by summarizing
our main findings.

%
%
\section{Preliminaries and Conventions}
\label{2}

Throughout this paper we work in the $\ov{\mbox{MS}}$--scheme
using an arbitrary
$R_\xi$--gauge for the gluon propagator
and the 't Hooft--Feynman gauge for the W--propagator.
Infrared (mass) singularities
are regulated by quark masses. Encouraged by \cite{bw} and \cite{k} we
use an anticommuting $\gamma_5$ (NDR scheme).

The algorithm to perform NLO QCD corrections to weak processes has
been outlined in \cite{bw,bjlw,bjw}, so that we just have to summarize
briefly the  necessary  steps with emphasis on the new features
in our calculation.

The part of $H^{\Delta S=2}$ in (\ref{s2})
relevant for $\eta_1$ reads
\begin{eqnarray}
H^c (\mu)&=&  \frac{G_F^2}{16 \pi^2} \mw  \lambda_c^2 \cdot
                 \eta_1 (\mu_W,\mu _b, \mu_c)
         \cdot     S( \frac{\mc(\mu_c)}{\mw} ) \cdot b(\mu) \cdot \oll(\mu ),
          \label{hcc}
\end{eqnarray}
where we have stressed  that  the calculation of $\eta_1$
involves the scales
$\mu _W \simeq M_W$, $\mu _b\simeq m_b$ and $\mu _c \simeq m_c$,
at which the W--boson, the bottom and the  charm quark are removed from the
theory.
A physical quantity cannot depend on these
scales
and any residual dependence
of $H^c$ on
them  is
due to the truncation of the perturbation series.
To judge the `theoretical
error' associated with this
we have varied   these scales instead of setting them exactly equal
to the values of
the corresponding current masses.
We will see that the NLO calculation  reduces the dependence of the LL
result on $\mu_W$, $\mu_b$ and $\mu_c$.

\subsection{The Hamiltonian at the Scale $\mu_W \simeq M_W$}\label{2thew}
\begin{figure}[htb]
\centerline{
\rotate[r]{
\epsfysize=14cm
\epsffile{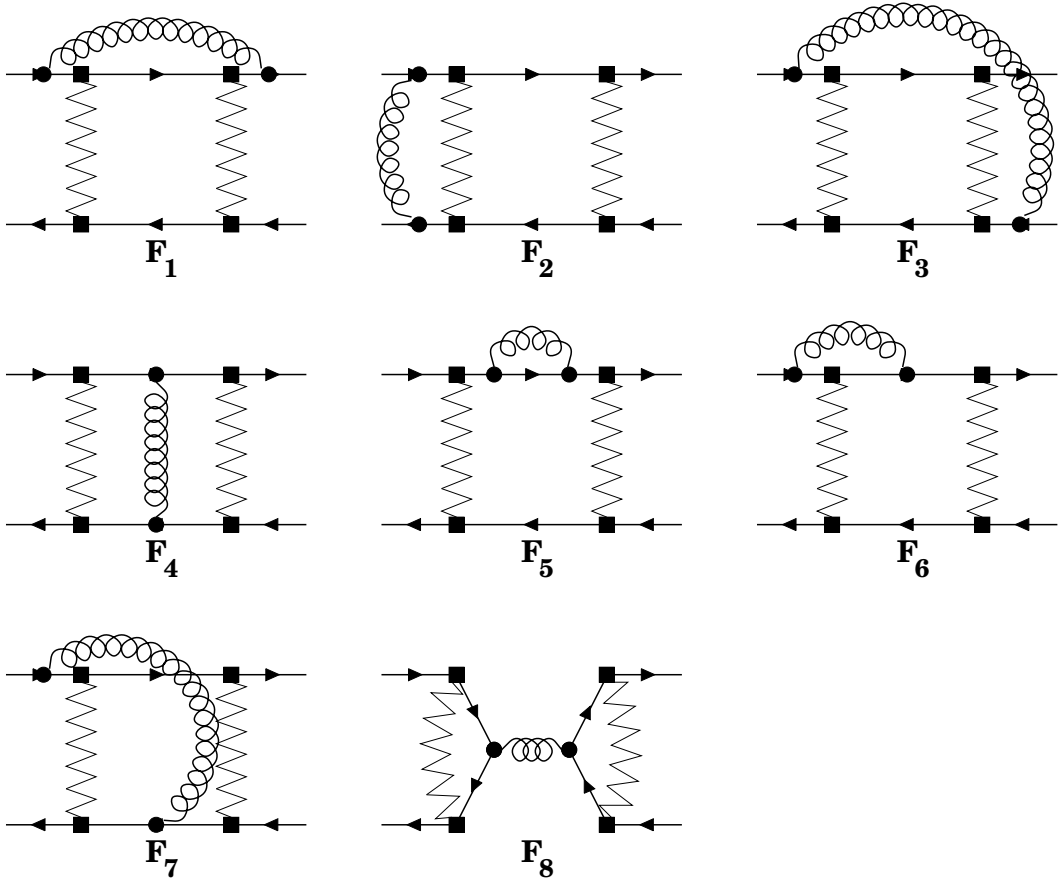}
}}
\caption[]{
	The classes of diagrams contributing to the $O(\alpha)$--correction
	to $\langle H^c \rangle $;
	the remaining diagrams are obtained by left--right and
	up--down reflections.
        The curly lines denote gluons.
	Also QCD counterterm diagrams have to be included.
	Diagram ${\rm F}_8$ evaluates to 0 for zero external momenta.
}
\label{boxqcd}
\end{figure}

Consider first the Green's function with the insertion  of four weak
currents calculated in the
full standard model.
The lowest order diagrams contributing to it  are shown in \fig{box},
the $O(\alpha)$--corrections are displayed in \fig{boxqcd}.

After multiplication by $i$  they constitute the
matrix element of $H^c$ between quark states
at the scale $\mu_W \simeq M_W$:
\begin{eqnarray}
\langle H^c (\mu_W) \rangle &=& \langle H(\mu_W)\rangle^{(0)}  +
 \frac{\alpha (\mu_W)}{4   \pi}  \langle H(\mu_W)\rangle^{(1)}
  +O(\alpha^2).
\end{eqnarray}
In view of the fact that the desired
short distance effects do
not depend on the treatment of the external states, one can choose
free quark states with zero external momenta
throughout the whole calculation.
The diagrams of \fig{box} with internal quark masses
$m_i,m_j$ involve
$\tilde{S}(m_i^2/\mw,m_j^2/\mw)$ \cite{il,bjw}, which enter
the coefficient of $\eta_1$ in the GIM--combination (with $m_u=0$)
\begin{eqnarray}
 \tilde{S}(\frac{\mc}{\mw},\frac{\mc}{\mw}) -
  2  \tilde{S}(\frac{\mc}{\mw},\frac{m_u^2}{\mw})
 +  \tilde{S}(\frac{m_u^2}{\mw},\frac{m_u^2}{\mw}) & =&
  S(\frac{\mc}{\mw})  \nn
& = &  \frac{\mc}{\mw} + O \lt( \frac{m_c^4}{M_W^4} \r)   \label{ilgim}
\end{eqnarray}
to give  \cite{gw}:
\begin{eqnarray}
\langle H(\mu_W)\rangle^{(0)} &=& \frac{G_F^2}{16 \pi^2}
                \lambda_c^2 \mc (\mu_W ) \langle \oll \rangle^{(0)}
.\label{h0}
\end{eqnarray}
The diagrams of \fig{boxqcd} have been calculated in \cite{bjw}
giving after inclusion of the GIM mechanism:
\begin{eqnarray}
\langle H(\mu_W)\rangle^{(1)} \! \! &=& \! \! \frac{G_F^2}{16 \pi^2}
                \lambda_c^2 \mc (\mu_W ) \lt\{ \!
          \langle \oll \rangle ^{(0)} h_{LL} \lt( \mu_W \r)+
      \langle \hat{T} \rangle ^{(0)} h_T  +
       \langle \hat{U} \rangle ^{(0)} h_U  \! \r\}.  \label{h1}
\end{eqnarray}
Only the lowest order in $\mc/\mw \ll 1$ has to be kept in (\ref{h0})
and (\ref{h1}).

In (\ref{h1}) new operators have emerged:
\begin{eqnarray}
\hat{T}&=& ( L \otimes L + R \otimes R  - \sigma _{\mu \nu}
         \otimes \sigma ^{\mu \nu} )\cdot \frac{N-1}{2 N} \1 , \nn
\hat{U} &=& \frac{1}{2} (  \gamma _\mu L \otimes \gamma ^\mu R +
               \gamma _\mu R \otimes \gamma ^\mu L ) \cdot
            \lt( \frac{N^2+N-1}{2 N} \1 - \frac{1}{2 N} \tw \r) \nn
&& - (  L \otimes R + R \otimes L ) \cdot
       \lt( \frac{N^2+N-1}{2 N} \tw - \frac{1}{2 N} \1 \r) , \label{u}
\end{eqnarray}
where $L=1-\gamma _5, R=1+\gamma _5$,
$N$ is the number of colours,
and $\1$ and $\tw$ denote colour
singlet and antisinglet, i.e. $(L \otimes R) \cdot \tw$ stands for
$\ov{s}_i (1-\gamma _5)  d_j \cdot
\ov{s}_j (1+\gamma _5)  d_i$.
We have written $\hat{U}$ in a manifestly Fierz self--conjugate way.
The Dirac structures of $\oll$ (see (\ref{oll})) and $\hat{T}$ are
also Fierz self--conjugate.

The
functions in (\ref{h1}) are:
\begin{eqnarray}
h_{LL} (\mu )\! &=& \! C_F \lt[- 1- 6 \log\frac{\mc}{\mu^2 }
  +  \xi \lt(2-2 \frac{\ms \log (\ms/\mu^2) -\md
                             \log (\md/\mu^2)}{\ms-\md}
               \r)        \r]     \nn
&& + \frac{N-1}{2 N} \lt[- 11+\frac{4}{3} \pi^2
                       + 12 \log\frac{\mc}{\mw}
                       +3  \log\frac{\md \ms}{\mu^4}
                       - 6 \log\frac{\mc }{\mu ^2} \r. \nn
 && \quad \hspace{1ex} \lt.   + \xi \lt( 2 + \log \frac{\md \ms}{\mu^4} -
                                   2 \frac{\ms \log(\ms/\mu^2) -
                                \md \log(\md/\mu^2) }{\ms-\md} \r)
                                  \r]  \nn
h_T  \! &=& \!  (-3 - \xi ) \nn
h_U  \! &=& \!
 \frac{3+\xi }{2} \frac{m_d m_s}{\ms-\md} \log \frac{\ms}{\md} ,\label{ltu}
\end{eqnarray}
where $C_F=(N^2-1)/(2 N)$.

Let us now discuss the structure of $\langle H(\mu_W) \rangle$ in
equations (\ref{h0}) to (\ref{ltu}):
Notice first that there is no large logarithm in (\ref{h0}),
the logarithms of the  individual box diagrams are cancelled by the
GIM mechanism.
In $h_{LL}$ in (\ref{ltu}) we have arranged
the logarithms such that one can easily distinguish
those which are small for $\mu \simeq m_c$
from the large logarithm
$\log( \mc /\mw )$. In the LL approximation
the terms of the form $ \lt[ \alpha /(4 \pi ) \cdot
\log( \mc /\mw ) \r]^n, n=0,1,2, \ldots $ are summed to all orders
in perturbation theory.

Next one observes that
$\langle H(\mu_W) \rangle^{(1)} $
involves the additional operators
$\hat{U}$
and
$\hat{T}$, which
are artefacts of the use of small quark masses $m_d$ and $m_s$
to regulate the mass singularities.
Since the desired Wilson coefficient must neither depend on
the gauge parameter $\xi$ nor on
the infrared regulators,
$h_U,h_T$ and the gauge part of $h_{LL}$ must disappear from
it, thereby providing a check
of the factorization described in section \ref{2fac}.
Yet in (\ref{ltu}) one can see that these terms do not depend
on $M_W$, which sets the scale for the short distance physics.

A dimensional regularization of the mass singularities would have
resulted in $h_U=h_T=0$.

\subsection{Factorization of the Hamiltonian at
            the Scale $\mu_W$}\label{2fac}
In order
to use the renormalization group (RG) evolution
to sum  the large logarithm $\log \lt( \mc/\mw \r)$
to all orders in perturbation theory  we seek a factorization
of the form
\begin{eqnarray}
H^c &\propto & \sum_{j} C_j  {\cal O}_j  \label{facw}
\end{eqnarray}
with composite operators ${\cal O}_j$ describing the weak
interaction in an effective field theory without
W-boson and top quark. The  short distance
physics is contained in the Wilson coefficients $C_j$,
to which renormalization
group methods will be applied.

Consider first the time ordered product of two weak currents
$J_{\mu} (x)$
folded with the W--propagator ${\cal D}^{\mu \nu}$, which mediates
a $\Delta S=1$ transition. Its operator product expansion is
well known to order  $\alpha$  \cite{acmp,bw,bjlw}:
\begin{eqnarray}
-i \int d^4 y \T J_{\mu}^{s j} (x) J_{\nu}^{k d} (y)
    {\cal D}^{\mu \nu} (x-y) &=&
 C_+ Q_+ ^{j k} (x) +C_- Q_- ^{j k} (x) +\ldots  \label{ope}
\end{eqnarray}
Here $j,k$ are flavour indices standing for  $u$ or $c$.
\begin{eqnarray}
Q_\pm^{jk} &=&  ( \ov{s} j )_{V-A}
   (\ov{k} d )_{V-A}  \cdot \frac{1}{2} \lt( \1 \pm \tw \r)
 \no
\end{eqnarray}
are dimension--six  four--quark
current-current operators, the meaning of $\1$ and $\tw$ has been
explained after eq.\  (\ref{u}).
The dots in (\ref{ope}) denote penguin operators,
which contribute to $\eta_3$ rather than $\eta_1$,
 and operators with dimension
higher than six.

Now the Green's function described by the diagrams of \fig{box}
and \ref{boxqcd} contains two
insertions  of (\ref{ope}) and thus  the effective
theory involves matrix elements with two insertions of $Q_+$
or $Q_-$. In order to find the complete set of operators
needed in (\ref{facw}) we have to look at the forthcoming
renormalization group analysis now:
The reader may have noticed that we have already put the
current--current operators into the basis
$(Q_+,Q_-)$, which is diagonal with respect to the renormalization
group evolution.
Further
the GIM mechanism prevents the mixing under renormalization
with penguin operators \cite{bjlw}.

\begin{figure}[htb]
\centerline{
\rotate[r]{
\epsfysize=5cm
\epsffile{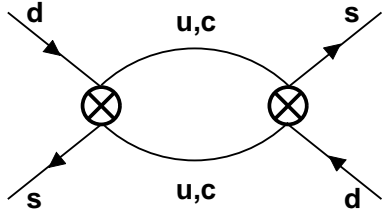}
}}
\caption[]{Diagram ${\rm D}_0$ in the effective four--quark theory
	constituting the matrix element of ${\cal O}_{ij}$ to
	order $\alpha^0$.
	The cross denotes the insertion of an effective
	$\Delta S = 1$ operator.
	}
\label{bi}
\end{figure}

\begin{figure}[htb]
\centerline{
\rotate[r]{
\epsfysize=14cm
\epsffile{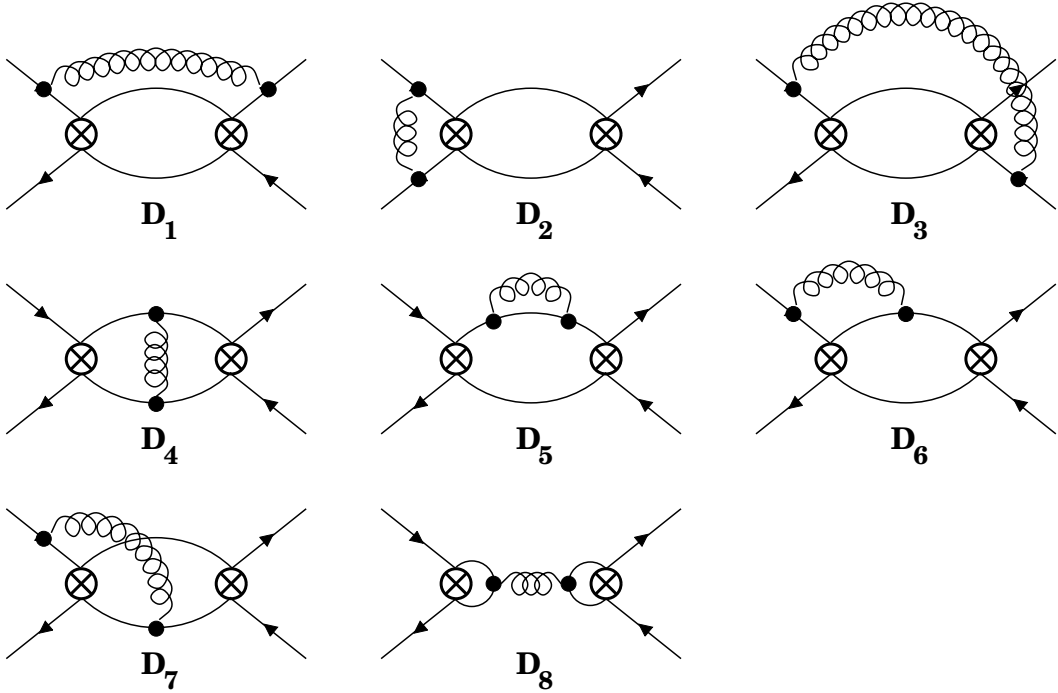}
}}
\caption[]{
	The classes of diagrams in the effective four--quark theory
        contributing to $O(\alpha)$.
	The other members of a given class are obtained by
	left--right and up--down reflections.
	Also QCD counterterms have to be included.
	Diagram ${\rm D}_8 = 0$ for zero external momenta.
}
\label{biqcd}
\end{figure}

For brevity we define  the bilocal operators
\begin{eqnarray}
{\cal O}_{ij}(x)  &=&  \frac{-i}{2} \int d^4 y \T
      \lt[  Q_i^{cc} (x) Q_j^{cc} (y) +Q_i^{uu} (x) Q_j^{uu} (y) \r. \nn
&& \lt. \quad \quad \quad \quad
   -  Q_i^{uc} (x) Q_j^{cu} (y) -Q_i^{cu} (x) Q_j^{uc} (y) \r]
     , \quad i,j=+,- \; .
\end{eqnarray}
The  diagrams contributing to the matrix element
$\langle {\cal O}_{ij} \rangle $
are depicted in \fig{bi} and \ref{biqcd},
the cross denotes the insertion of  $Q_i$  or $Q_j$.

Now one expects the diagrams of \fig{bi} and \ref{biqcd}
to be divergent enforcing an extra renormalization of
${\cal O}_{ij}$ proportional to a local four--quark operator.
The GIM mechanism, however, cancels the divergences.
Clearly in \fig{biqcd} one--loop QCD counterterms
must be included first.
So we end up with
\begin{eqnarray}
H^c (\mu) &=& \frac{G_F^2}{2} \lambda_c^2
\sum_{i,j=+,-} C_{ij} (\mu)  {\cal O}_{ij} (\mu) .  \label{matchw}
\end{eqnarray}

These features have been discovered by Witten \cite{w} a long time
before we have carried out this explicit calculation.
He has considered four--quark QCD and states:
\begin{eqnarray}
H^{c} (x) = \frac{-i}{2} \int d^4 y \T \lt[
   H^{c,\Delta S=1} (x)   H^{c,\Delta S=1} (y) \r] , \no
\end{eqnarray}
where  we have denoted the part of the $\Delta S=1$ hamiltonian
containing the first two quark generations by
$H^{c,\Delta S=1}$.
For our case this implies that we can simply obtain  $C_{ij}$
by multiplying the $\Delta S=1$ Wilson--coefficients $C_i$ and
$C_j$ calculated in \cite{bw}, where the $\Delta S=1 $ hamiltonian
is normalized as
\begin{eqnarray}
 H^{\Delta S=1}= \frac{G_F}{\sqrt{2}}
	\sum_{i=+,-}
	\sum_{kl}
	V_{ks}^*  V_{ld}\; C_{i}\; Q_{i}^{kl} . \no
\end{eqnarray}
This multiplicativity of Wilson coefficients has also
been proven for a scalar theory by Lee \cite{l}.
Since we anyway need the finite parts of the
diagrams of \fig{bi} and \ref{biqcd}
later in the calculation (section \ref{2the}), we
have compared $\sum C_i (\mu_W) C_j (\mu_W )
\langle {\cal O}_{ij} (\mu_W) \rangle$
with $\langle H^c (\mu _W)\rangle $ as given in eq.\ (\ref{h0}) and
(\ref{h1}) and have indeed verified
\begin{eqnarray}
C_{ij}(\mu_W)&=&C_i(\mu_W) C_j(\mu _W) \label{witten}
\end{eqnarray}
to order $\alpha$.

\subsection{Evolution of the
            Hamiltonian from $\mu_W$ to $\mu_c \simeq m_c$}\label{2evo}
In  the next step we want to evolve $C_{ij} (\mu)$ from $\mu _W$ down
to $\mu_c$ using the renormalization group formalism.
This procedure is well known \cite{bw,bjw,bjlw} and we briefly list
its main ingredients:

The running of the coupling constant $g$
($\alpha = g^2/(4 \pi) $)
is determined by the
beta function
\begin{eqnarray}
\beta (g) &=& -\beta _0^{(f)}  \frac{g^3}{16 \pi^2} -\beta _1^{(f)}
              \frac{g^5}{\lt( 16 \pi^2 \r)^2} + O(g^7) \label{betaseries}
\end{eqnarray}
with $f$ being the number of flavours and
\begin{eqnarray}
\beta _0^{(f)} &=& \frac{11 N -2 f}{3}, \quad
 \beta _1^{(f)} \; = \; \frac{34}{3} N^2 - \frac{10}{3} N f -2 C_F f
\label{beta}.
\end{eqnarray}
$\alpha (\mu)$ reads  in terms of the f--flavour
QCD scale parameter $\Lambda_f$:
\begin{eqnarray}
\frac{\alpha ( \mu  ) }{4 \pi } &=&
       \frac{1}{\beta_0^{(f)}\log(\mu^2/\Lambda_f^2)} -
                \frac{\beta_1^{(f)} \log \lt[ \log(\mu^2/\Lambda_f^2) \r]}{
                 \lt( \beta_0^{(f)} \r)^3 \log ^2 (\mu^2/\Lambda_f^2)}
+ O \lt( \frac{ \log^2[\ldots] }{
                 \log^3 (\ldots ) } \r) . \label{runalpha}
\end{eqnarray}
$\Lambda_3$ and $\Lambda_5$ are obtained from
$\Lambda_4=\Lambda_{\ov{\rm MS}}$ by imposing continuity on $\alpha$.

The anomalous dimension $\gamma _{ij}$ of ${\cal O}_{ij}$ are readily
obtained from those of $Q_\pm$  via their definition
in terms of the renormalization constants $Z_\pm$.
Since no matrix renormalization is required for $Q_\pm$ and the
bilocal operators
${\cal O}_{ij}$ do  not mix with local operators,
also
${\cal O}_{ij}$ renormalizes multiplicatively, so that
\begin{eqnarray}
\gamma _{ij} &=& \lt( Z_{i} Z_j \r)^{-1} \mu \frac{d}{d \mu }
                  \lt( Z_{i} Z_j \r)
\; = \; \gamma _i +\gamma _j, \quad i,j=+,-\no .
\end{eqnarray}
The coefficients $\gamma _\pm^{(0)}$ and  $\gamma _\pm^{(1)(f)}$ of
the anomalous dimension of $Q_\pm$,
\begin{eqnarray}
\gamma _\pm (g)&=&  \gamma _\pm^{(0)} \frac{g^2}{16 \pi^2} +
                    \gamma _\pm^{(1)(f)} \lt( \frac{g^2}{16 \pi^2} \r)^2
                    + O(g ^6), \no
\end{eqnarray}
read in the NDR scheme \cite{bw}:
\begin{eqnarray}
\gamma _\pm^{(0)}&=& \pm 6 \frac{N \mp 1}{N}, \quad
\gamma _\pm^{(1)(f)} \; = \;   \frac{N \mp 1}{2 N}
    \lt(-21 \pm \frac{57}{N} \mp \frac{19}{3} N \pm \frac{4}{3} f  \r)
\label{gammapm} .
\end{eqnarray}
We define for $i,j=+,-$:
\begin{eqnarray}
d_{j}^{(f)} &=& \frac{\gamma _{j}^{(0)}}{2 \beta _0^{(f)}} ,
\quad \quad \quad  d_{ij}^{(f)}
\; = \; d_i^{(f)} + d_j^{(f)} ,\nn
Z_{j}^{(f)} & = &
\frac{\gamma _{j}^{(1)} -2 \beta _1^{(f)} d_{j}^{(f)} }{2 \beta_0^{(f)} },
\quad \quad \quad
Z_{ij}^{(f)}
\; = \; Z_{i}^{(f)} + Z_{j}^{(f)}
\label{dz} ,
\end{eqnarray}
so that the solution to the renormalization group equation
for $C_{ij} (\mu )$
reads 
below the b--quark threshold $\mu_b$ in NLO:
\begin{eqnarray}
C_{ij} (\mu )&=&
 \! \! C_{ij} (\mu _W)
           \lt( \frac{\alpha (\mu _W)}{\alpha (\mu _b)}   \r) ^{d_{ij}^{(5)}}
           \lt( \frac{\alpha (\mu _b)}{\alpha (\mu )}   \r) ^{d_{ij}^{(4)}}
 \cdot    \nn
&& \quad \lt( 1 + \frac{\alpha (\mu _W)- \alpha (\mu _b)}{4 \pi } Z_{ij}^{(5)}
   + \frac{\alpha (\mu _b) - \alpha (\mu )}{4 \pi } Z_{ij}^{(4)}
             \r) \nn
&=&  C_{i} (\mu ) C_{j} (\mu ). \label{wittenren}
\end{eqnarray}
Hence the RG improved Wilson coefficient $C_{ij}$ also equals $C_i \cdot C_j$.

\subsection{The Hamiltonian below $\mu_c$}\label{2the}
Below the scale $\mu _c \simeq m_c$ the c--quark field is no more a
dynamic degree of freedom and we have to express $H^c (\mu )$ in an
effective three--quark theory.
Here we are left with the  single local four--quark operator $\oll$ defined
in eq.\ (\ref{oll}).

\begin{figure}[hbt]
\centerline{
\rotate[r]{
\epsfysize=3cm
\epsffile{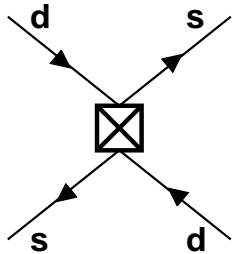}
}}
\caption[]{
	The diagram for the matrix element of $\oll$
	in the effective three--quark theory to order
	$\alpha^0$.
	The cross denotes the insertion of the effective
	$\Delta S = 2$ operator.
	}
\label{loc}
\end{figure}

\begin{figure}[hbt]
\centerline{
\rotate[r]{
\epsfysize=14cm
\epsffile{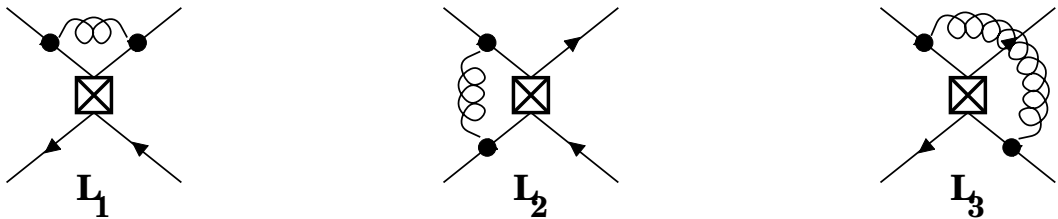}
}}
\caption[]{
	Classes of diagrams in the effective three--quark theory
	contributing to $\langle \oll \rangle$ to order $\alpha$.
	The other members of a given class are obtained by
	left--right and up--down reflections.
	QCD counterterms have to be included.
	}
\label{locqcd}
\end{figure}

To obtain the Wilson coefficient $C$ in
\begin{eqnarray}
H^c (\mu ) &=& \frac{G_F^2}{16 \pi ^2} \lambda _c^2 \; C (\mu ) \oll (\mu)
\label{hmc}
\end{eqnarray}
one must first solve the matching condition between the four-- and
three--quark theory:
\begin{eqnarray}
\frac{G_F^2}{2} \lambda _c^2 \sum_{i,j=+,-} C_{ij} (\mu _c)
     \langle {\cal O}_{ij} (\mu_c ) \rangle
&=& \frac{G_F^2}{16 \pi ^2 } \lambda _c^2 \; C (\mu _c)
    \langle \oll (\mu _c) \rangle
    . \label{matchc}
\end{eqnarray}
At this stage the matrix elements $\langle {\cal O}_{ij} \rangle$ must be
calculated.
$\langle \oll \rangle$ has already been obtained
to order $\alpha$ in \cite{bjw}.
The corresponding diagrams are depicted in
\fig{loc} and \ref{locqcd}.

Since the LL analysis sums up
the terms
$\lt[ \alpha \cdot \mbox{large log}\r]^n$
for  $n=0,1,2,\ldots,$
to all orders
in perturbation theory,
the LL matching has to be made to order
$\alpha ^0$.
Consequently in NLO
the terms
$\alpha \lt[ \alpha \cdot \mbox{large log}\r]^n$, $n=0,1,2,\ldots,$
are summed,
and
one has to match the effective theories to order
$\alpha ^1$.
For this
the finite parts of
the diagrams of \fig{biqcd}
must be calculated.

This is in contrast to the recent NLO analysis of
$K^+ \rightarrow \pi ^+ \nu \ov{\nu }$,
$K_L \rightarrow \mu ^+ \mu ^-$ by  Buchalla and  Buras
\cite{bb}, which also involves the renormalization of bilocal operators.
There the one--loop Green's function
already contains
a large logarithm and therefore the  NLO matching
can be made at the one--loop level.

The evolution of $C(\mu )$ in (\ref{hmc}) to scales below $\mu _c $ is
dictated by the anomalous dimension of $\oll$, which equals $\gamma_+$
in (\ref{gammapm}) \cite{bw}:
\begin{eqnarray}
C(\mu ) &=& C (\mu _c) \lt( \frac{ \alpha (\mu _c)}{\alpha (\mu) } \r)
                        ^{d_+^{(3)} }
    \lt( 1+ \frac{\alpha (\mu _c)-
  \alpha (\mu )}{4 \pi } Z_+^{(3)} + O( \alpha ^2) \r)
  \label{belowc} .
\end{eqnarray}
The residual $\mu $--dependence in (\ref{belowc}) is absorbed into
$b(\mu )$ defined in (\ref{s2}):
\begin{eqnarray}
b(\mu ) &=& \lt[ \alpha (\mu)   \r] ^{-d_+^{(3)}}
            \lt( 1 - \frac{\alpha (\mu )}{4 \pi } Z_+^{(3)}
       + O( \alpha ^2) \r) . \label{smallb}
\end{eqnarray}

We close this section by noting that $C (\mu) $
has via (\ref{matchc})
absorbed all  dependence on the charm quark mass present in
$\langle {\cal O}_{ij} (\mu_c)   \rangle $,
because the
matrix element $\langle \oll (\mu_c) \rangle $ is evaluated
in an effective three-quark
theory containing no information on $m_c$. Consequently
$C(\mu)$  contains the running charm quark mass  in the form $m_c (\mu_c)$.

To discuss the dependence of our result on the scale
$\mu _c$ in section  \ref{5}, we need the anomalous mass dimension
$\gamma _m$ , which is given by
\begin{eqnarray}
\gamma _m (g) &=& \gamma _m^{(0)} \frac{g^2}{16 \pi^2} +
              \gamma _m^{(1)(f)} \lt( \frac{g^2}{16 \pi^2} \r) ^2 +
              O \lt( g ^6  \r) \no
\end{eqnarray}
with
\begin{eqnarray}
\gamma_m^{(0)}&=& 6 C_F , \quad
\gamma _m^{(1)(f)} \; = \; C_F \lt( 3 C_F +
                  \frac{97}{3} N - \frac{10}{3} f \r)
\label{gammam}.
\end{eqnarray}
Hence with
\begin{eqnarray}
d_m^{(f)}  &=& \frac{\gamma_m^{(0)}}{2 \beta_0^{(f)}} \quad \mbox{and } \quad
Z_m^{(f)} \; = \;
  \frac{\gamma_m^{(1)(f)}-2 \beta_1^{(f)} d_m^{(f)}}{2 \beta_0^{(f)}} \no
\end{eqnarray}
the running mass reads
\begin{eqnarray}
m_c (\mu_c)&=& m_c ( m_c )
\lt( \frac{\alpha (\mu_c)}{\alpha (m_c)} \r)^{d_m^{(f)}}
\lt( 1+ \frac{\alpha (\mu_c) - \alpha (m_c)}{4 \pi} Z_m^{(f)} \r).  \label{rum}
\end{eqnarray}

%
%
\section{The Leading Log Analysis}
\label{3}

In this section we review the LL analysis of \cite{gw}
using the formalism of
section \ref{2} and taking into account that the top quark is heavy.
This analysis has already been performed by Flynn \cite{f}.
A good approximation neglecting quark thresholds has been given by
Datta, Fr\"ohlich and Paschos \cite{dfp}.
We discuss the scale dependences inherent to the LL calculation.
The relevant RG formulae are obtained from those in section
\ref{2} by setting $\beta_1^{(f)}$,
$\gamma_{\pm}^{(1) (f)}$
and
$\gamma_{m}^{(1) (f)}$ to zero.

\subsection{The Hamiltonian between the Scales $\mu_c$ and $\mu_W$}
\label{3w}

At first we want to  factorize  the $H^c (\mu)$  for $\mu_c \leq \mu
\leq \mu_W$
in order to separate short and  long distance physics
as described
in sections \ref{2fac} and \ref{2evo}:
\begin{eqnarray}
\langle  H^c (\mu ) \rangle &=&\frac{G_F^2}{2} \lambda_c^2
  \sum _{i,j=+,-} C_{ij} (\mu )
\langle {\cal O}_{ij} (\mu )\rangle^{(0)} . \label{llfacw}
\end{eqnarray}
At the matching scale $\mu_W$
the left hand side of (\ref{llfacw}) simply equals
$\langle H^c (\mu_W)\rangle ^{(0)}$
given in (\ref{h0}).
The $\Delta S=1$ Wilson coefficients $C_\pm( \mu_W)$
are equal to 1 in the LL
approximation \cite{bw,gw2}, so that we get
$C_{ij} ( \mu_W)=1$ by means of
(\ref{witten}).


In section \ref{3c} we will need (see \fig{bi})
\begin{eqnarray}
\langle {\cal O}_{ij} (\mu) \rangle^{(0)}&=&
      \tau_{ij} \frac{ \mc (\mu ) }{8 \pi^ 2}
                   \langle \oll \rangle^{(0)}  \label{tau}
\end{eqnarray}
with
\begin{eqnarray}
\tau_{++} &=& \frac{N+3}{4}, \quad
\tau_{+-} \; = \; \tau_{-+} \; = \; \frac{-N+1}{4}, \quad
\tau_{--} \; = \;  \frac{N-1}{4}.  \no
\end{eqnarray}
Using (\ref{tau})
and (\ref{h0}) one easily checks (\ref{llfacw}).

$\langle H^c ( \mu_W) \rangle$  contains the Inami--Lim
function $S$, which has been expanded to lowest order in $\mc / \mw$ in
(\ref{ilgim}). To include  higher order terms, $\lt( \mc / \mw \r)^2 \ldots$
in the factorization (\ref{llfacw}), one must take into account
 operators with
dimension higher than six in (\ref{ope}).

Next we seek the  resummation of the large logarithm
$\alpha \log \lt( \mc/\mw \r) $
present in the  higher order terms of
$H^c$.
The factorization  (\ref{llfacw}) splits  $\log \lt( \mc/\mw \r) $
into $  \log \lt( \mu^2/\mw \r) +\log \lt( \mc/\mu^2 \r)  $.
While the latter resides  in the
matrix elements,
 the former
is contained in the Wilson coefficients and is summed to all orders
by the
renormalization group evolution from $\mu_W$ to $\mu_c$.

The LL running of the Wilson coefficients down to $\mu_c$ is
given by
\begin{eqnarray}
C_{ij} (\mu_c )&=&
 \! \!      \lt( \frac{\alpha (\mu _b)}{\alpha (\mu_c )}   \r) ^{d_{ij}^{(4)}}
           \lt( \frac{\alpha (\mu _W)}{\alpha (\mu _b)}   \r) ^{d_{ij}^{(5)}}
            C_{ij} (\mu _W)
, \label{cll}
\end{eqnarray}
where $d_{ij}$ has been defined in (\ref{dz}), and the LL running for
$\alpha$ must be used.

Now we show the effect of   the  RG resummation on (\ref{llfacw}).
Setting $\mu_W=M_W$, $\mu_c=m_c$ for the moment and expanding
(\ref{cll})
around $M_W$ to order $\alpha \log \lt( \mc/\mw \r)$  yields
with (\ref{tau}):
\begin{eqnarray}
\langle  H^c (m_c ) \rangle
&=&         \frac{G_F^2}{16 \pi^2}
                \lambda_c^2 \mc (m_c ) \langle \oll \rangle^{(0)}
              \cdot \nn
&&           \lt[ 1 + \frac{\alpha }{4 \pi} \log \lt( \frac{\mc}{\mw} \r)
          \sum_{i,j=+,-} \frac{ \gamma_{ij}^{(0)} }{2} \tau_{ij}
+O\lt( \alpha^2 \log^2 \lt(\frac{\mc}{\mw} \r) \r) \r] .\label{expandw}
\end{eqnarray}
Since $\sum \tau _{ij} \gamma_{ij}^{(0)}/2 = 12 (N-1)/(2 N) $, we
have reproduced in (\ref{expandw}) the large logarithm present in
the coefficient $h_{LL}$ of (\ref{h1}).

\subsection{The Hamiltonian below the Scale $\mu_c$}\label{3c}
At the scale $\mu_c$ we obtain $C(\mu_c)$ from the matching
condition (\ref{matchc}):
\begin{eqnarray}
C(\mu_c) &=& \mc (\mu_c) \sum_{i,j=+,-} \tau_{ij}
         \lt( \frac{\alpha (\mu_b)}{\alpha (\mu_c)}\r)^{d_{ij}^{(4)}}
       \lt( \frac{\alpha (\mu_W)}{\alpha (\mu_b)}\r)^{d_{ij}^{(5)}}.
\end{eqnarray}
The LL evolution of $C(\mu)$  below $\mu_c$
can be obtained from the corresponding NLO expression in (\ref{belowc})
by setting $Z_+^{(3)}$ to zero.

Hence $\eta_1$ as defined in (\ref{s2}) or (\ref{hcc})
reads in the LL approximation:
\begin{eqnarray}
\eta_1^{\rm LO}\! \! &=&
         \! \! \lt( \alpha (\mu_c)  \r)^{d_+^{(3)}} \sum_{i,j=+,-} \tau_{ij}
         \lt(\frac{\alpha (\mu_b)}{\alpha (\mu_c)} \r)^{d_{ij}^{(4)}}
         \lt(\frac{\alpha (\mu_W)}{\alpha (\mu_b)} \r)^{d_{ij}^{(5)}}
           \nn
&=&\! \! \lt( \alpha (\mu_c )  \r)^{2/9} \! \lt[ \frac{3}{2}
         \lt(\frac{\alpha (\mu_b)}{\alpha (\mu_c)} \r)^{12/25}  \!
     \lt(\frac{\alpha (\mu_W)}{\alpha (\mu_b)} \r)^{12/23}  \!
 \!     \! \! \!
-         \lt(\frac{\alpha (\mu_b)}{\alpha (\mu_c)} \r)^{-6/25} \!
    \lt(\frac{\alpha (\mu_W)}{\alpha (\mu_b)} \r)^{-6/23}  \!
 \!     \! \! \! \! \! \r. \nn
&& \lt. \quad \quad \quad \quad
+ \frac{1}{2}
         \lt(\frac{\alpha (\mu_b)}{\alpha (\mu_c)} \r)^{-24/25} \!
  \lt(\frac{\alpha (\mu_W)}{\alpha (\mu_b)} \r)^{-24/23}    \r] .
\label{etall}
\end{eqnarray}

In the preceding sketch of the LL calculation we have seen how the
various scales $\mu_W$, $\mu_b$ and $\mu_c$  enter the result of
$\eta_1^{\rm LO}$. $H^c$, which is related to physical observables,
must not depend on these scales, and we can vary them
in (\ref{etall}) to judge the
accuracy of the perturbative calculation.
The residual scale dependence of (\ref{etall}) is sizeable and is
substantially reduced in the NLO result. A comparative numerical analysis
of the LL and NLO scale setting ambiguity will be postponed until section
\ref{5}, and at this place we just display the sources of the scale
dependence analytically.

The coupling constant grows with decreasing  scale, $H^c$ is
therefore most sensitive to the variation of the smallest scale
$\mu_c$, and we exemplify the scale dependence by expanding
the  concerning parts of the hamiltonian around
$m_c^*  = m_c (m_c) $. They read
\begin{eqnarray}
\eta_1^{\rm LO} (\mu_c) \mc (\mu_c)  \!
&=&\!
  \lt( m_c^* \r)^2 \lt( \alpha (m_c^*) \r) ^{d_+^{(3)}}
       \sum_{i,j=+,-}
           \lt(\frac{\alpha (\mu_b)}{\alpha (m_c^*)}\r)^{d_{ij}^{(4)}}
             \lt(\frac{\alpha (\mu_W)}{\alpha (\mu_b)} \r)^{d_{ij}^{(5)}}
\cdot \nn
&& \! \! \! \! \! \! \! \! \! \! \! \tau_{ij}
  \lt[ 1 + \frac{\alpha}{4 \pi} \log  \frac{\mu_c^2 }{\lt( m_c^*\r)^2}
     \lt( \frac{\gamma_{ij}^{(0)} -\gamma_+^{(0)}}{2}
               - \gamma_m ^{(0)}    \r)
      + O\lt( \alpha^2 \log^2 \frac{\mu_c^2 }{\lt( m_c^*\r)^2} \r) \r]
       \! .
\label{explog}
\end{eqnarray}
Notice that the largest contribution to the coefficient of
$\alpha \log (\mu_c^2/m_c^2)$ comes from the anomalous dimension of
 ${\cal O}_{--}$.
The two--loop finite part of the NLO analysis  will encounter a term
which  cancels the logarithm in (\ref{explog}) and will thereby diminish
the scale ambiguity.

%
%
\section{The Next--to--Leading Order Analysis}
\label{4}

In this section we describe in detail our NLO calculation, in which
the $O(\alpha)$--contribution to the transition amplitude (\fig{boxqcd})
is improved by summing the terms
$\alpha ^{n+1}  \log^{n} \lt(  \mc/\mw   \r)$, $n=0,1,2,\ldots,$ to all
orders in perturbation theory.
\subsection{The Hamiltonian between the Scales $\mu_c$ and $\mu_W$}\label{4w}
For the NLO factorization
\begin{eqnarray}
\langle  H^c (\mu ) \rangle  &=&\frac{G_F^2}{2} \lambda_c^2
  \sum _{i,j=+,-} C_{ij} (\mu )
\langle {\cal O}_{ij} (\mu )    \rangle, \quad
 \mu_c \leq \mu \leq \mu_W ,\label{nlofacw}
\end{eqnarray}
we need the $\Delta S=1$ Wilson coefficients $C_{\pm}$
to order $\alpha$. At the matching point one has \cite{bw}:
\begin{eqnarray}
C_\pm ( \mu_W ) &=& 1 + \frac{\alpha (\mu_W) }{4 \pi}
     \lt( \frac{\gamma_\pm^{(0)}}{2} \log \frac{\mu_W ^2}{\mw}
          + B_{\pm} \r) +O (\alpha^2) \label{wc1}
\end{eqnarray}
with $B_{\pm}$ given in the NDR scheme
\begin{eqnarray}
B_\pm &=&  \pm 11 \frac{N \mp 1}{2 N}.   \label{bpm}
\end{eqnarray}
Hence in (\ref{nlofacw})
\begin{eqnarray}
C_{ij} ( \mu_W   ) &=& 1 + \frac{\alpha (\mu _W ) }{4 \pi}
     \lt( \frac{\gamma_{ij}^{(0)}}{2} \log \frac{\mu_W   ^2}{\mw}
          + B_i  + B_j \r) + O\lt(\alpha^2 \r). \label{wc}
\end{eqnarray}

As outlined in section \ref{2thew}  the
factorization  (\ref{nlofacw}) is so simple because of the absence
of mixing between ${\cal O}_{ij} $ and local operators.
Let us briefly explain why they do not mix: The diagrams of \fig{bi}
and \ref{biqcd} contain two internal quarks with masses
$m_k$ and $m_l$, each of which is $m_u$ or $m_c$.
The elements of the anomalous dimension matrix
responsible for the mixing
can directly  be read
off from the divergent parts of these diagrams. Since the
diagrams have dimension 2 and the
anomalous
dimension  does not depend explicitly on $\mu$ in a mass independent
renormalization scheme as $\ov{\mbox{MS}}$, the singular part of
the diagrams must be proportional to $m_k^2 +m_l^2$ by symmetry under
$m_k \leftrightarrow m_l$. Hence  in the
GIM--combination $(k,l)=(c,c)-2 (c,u)+(u,u)$
the divergent parts cancel and
no mixing with local operators occurs.

Since we had to calculate the two--loop contributions
to $\langle {\cal O}_{ij}     \rangle$
shown in \fig{biqcd} for the matching at $\mu_c$
described in section \ref{4c},
we could  verify (\ref{nlofacw})
and  analyze the contributions of the individual diagrams to
$C_{ij}$:
The diagrams ${\rm D}_2$, ${\rm D}_4$,
${\rm D}_6$ and ${\rm D}_7$ contain each one
tree--level and one
 dressed $\Delta S=1$ four--quark operator
as subdiagrams.
All other diagrams can be divided
by a three--particle cut through the gluon line and the $u/c$--quark lines
into two subdiagrams, each of which contains a tree--level
$\Delta S=1$ operator with four (anti-) quarks and
one gluon as external states.
In (\ref{nlofacw}) the latter  match their corresponding
counterparts of \fig{boxqcd} with the tree--level Wilson coefficient,
thus these
diagrams are also properly taken into account  in (\ref{nlofacw}).
 Hence there is no point in
treating the diagrams ${\rm D}_1$, ${\rm D}_3$, ${\rm D}_4$
and ${\rm D}_5$
differently
from the others as done by the authors of
\cite{fky}.
In fact only ${\rm D}_2$, ${\rm D}_4$
and ${\rm D}_7$ contribute to the $O(\alpha)$--part in
(\ref{wc1}). ${\rm D}_6$ contains a dressed colour singlet
$\Delta S=1$  operator, which does not contribute to the
$O(\alpha)$--part of $C_\pm$ \cite{bw}, and therefore matches the diagram
${\rm F}_6$ with the tree--level Wilson coefficient.
Finally the term with $\log (\mu_W^2/\mw) $ in (\ref{wc1})
originates solely from ${\rm D}_7$. We will see in a moment that this
term will reduce the dependence of $C_{ij}$ on $\mu_W$.

The RG improved Wilson coefficient at $\mu _c$ is immediately
obtained from (\ref{wittenren}) and (\ref{wc}):
\begin{eqnarray}
C_{ij} (\mu_c )&=&
 \! \!
           \lt( \frac{\alpha (\mu _b)}{\alpha (\mu_c )}   \r) ^{d_{ij}^{(4)}}
 \! \!     \lt( \frac{\alpha (\mu _W)}{\alpha (\mu _b)}   \r) ^{d_{ij}^{(5)}}
\lt[ 1
   + \frac{\alpha (\mu _b) - \alpha (\mu_c )}{4 \pi } Z_{ij}^{(4)}
   + \frac{\alpha (\mu _W)- \alpha (\mu _b)}{4 \pi } Z_{ij}^{(5)}
     \r.   \nn
&&
\hspace{4cm}
\lt. + \frac{\alpha (\mu _W) }{4 \pi }
  \lt( \frac{\gamma _{ij}^{(0)}}{2} \log \frac{\mu _W^2}{\mw} + B_i + B_j  \r)
            \r] . \label{wcc}
\end{eqnarray}
We close this section by noting that in (\ref{wcc})
the LL scale  ambiguity
at $\mu_W$ is removed to order $\alpha $ due to
\begin{eqnarray}
\lt[ \alpha (\mu _W ) \r]^{d_{ij}^{(5)}} \lt( 1+\frac{\alpha (\mu _W)}{4 \pi }
            \frac{\gamma _{ij}^{(0)} }{2}
        \log \frac{\mu _W ^2}{\mw}      \r)
&=& \lt[ \alpha (M _W ) \r]^{ d_{ij}^{(5)} }   + O( \alpha^2 ) . \no
\end{eqnarray}

\subsection{The Hamiltonian below the Scale $\mu_c$}\label{4c}
To obtain the hamiltonian for  $\mu \leq \mu_c$
given in (\ref{hmc}) we must first solve the matching
condition between the three-- and four--quark theory
for $C (\mu )$:
\begin{eqnarray}
\langle H^c (\mu _c) \rangle \; = \;
\frac{G_F^2}{2} \lambda _c^2 \sum_{i,j=+,-} C_{ij} (\mu _c)
     \langle {\cal O}_{ij} (\mu_c ) \rangle
&=& \frac{G_F^2}{16 \pi ^2 } \lambda _c^2 \; C (\mu _c)
    \langle \oll (\mu _c) \rangle
    . \label{4matchc}
\end{eqnarray}
As discussed in section \ref{2the} this step requires the calculation
of the finite parts of the  two--loop diagrams
shown in \fig{biqcd}.

\subsubsection{Calculation of the Two--Loop Diagrams in
	Fig.\ \protect{\ref{biqcd}}}
We now sketch some details of the two--loop calculation, which has
been carried out independently by both of us.

The diagrams of \fig{biqcd} in arbitrary $R_\xi$--gauge for the gluon
involve tensor integrals up to rank
six.
As explained in section \ref{2thew} we could set the external momenta to zero,
the masses $m_{k,l}$ of the internal quark lines connecting the four--quark
operators were kept arbitrary and in the very end set
equal to $m_c$ or to $m_u=0$.
Diagrams
${\rm D}_1$ through
${\rm D}_3$ are infrared singular and we have kept nonzero
strange and down quark masses for their regularization.
Alternatively one could  use an off--shell external momentum
or regulate both UV-- and IR--singularities dimensionally.
Needless to say that the same regularization must be used
for all matrix elements in (\ref{4matchc}).
It turned out to be convenient to calculate the integrals keeping the
space--time dimension $D$ arbitrary,
which has resulted in quite compact expressions for the coefficients
of the various tensors built out of combinations
of metric tensors. These coefficients involve
Gau\ss' hypergeometric function ${}_2 {\rm F}_1$,
whose expansion in terms of $\varepsilon =(4-D)/2$ yields quite lengthy
expressions containing the familiar dilogarithm function in the finite part.
${\rm D}_1$, ${\rm D}_2$ and  ${\rm D}_3$
also contain a additive terms with the
IR--regulators.
The dilogarithms
disappear
for $m_k = m_l=m_c$ or $m_{k,l} = m_u \rightarrow 0$.
Only a few number of the hypergeometric functions must be expanded explicitly,
the others can be obtained via functional equations.
As a byproduct we have obtained the results corresponding to a dimensional
infrared regularization.
The $1/\varepsilon ^2$--terms of the diagrams disappear in the
GIM--combination.

The $D$--dimensional Dirac algebra has been evaluated
with the help of the
package {\sc Tracer}  \cite{jl} for the
computer algebra system {\sc Mathematica}.
The  results
involve products of up to five
Dirac matrices on each fermion line. They have to be projected onto
a set of Dirac structures which form a basis for $D=4$.
We have chosen the following convention:
\begin{eqnarray}
\gamma _{\mu} \gamma _{\nu}  \gamma _{\vartheta } L \otimes
\gamma ^{\mu} \gamma ^{\nu}  \gamma ^{\vartheta } L &= &
(16 - 4 \varepsilon ) \:
    \gamma _{\nu} L \otimes \gamma ^{\nu} L  + \mbox{ev.}  \nn
\gamma_{\rho } \gamma_{\delta }
\gamma _{\mu} \gamma _{\nu}  \gamma _{\vartheta } L \otimes
\gamma^{\rho } \gamma^{\delta }
\gamma ^{\mu} \gamma ^{\nu}  \gamma ^{\vartheta } L &= &
(256 - 224 \varepsilon ) \:
    \gamma _{\nu} L \otimes \gamma ^{\nu} L  + \mbox{ev.}
\label{greek}
\end{eqnarray}
where $L=1-\gamma_5$ and
ev.\ stands for  evanescent operators vanishing for $D=4$.
The infrared singular diagrams contain additional Dirac structures,
which multiply only finite integrals.

The same procedure has to be performed for the
graphs with QCD counterterms for the  inserted $\Delta S=1$ operators.
The diagrams  of \fig{biqcd} also contain  divergent subloops
involving both $\Delta S=1$ operators, their divergent parts vanish in
the GIM--combination like the divergence of \fig{bi}, so that no
counterterms for these subdivergences are needed.

After summing the diagrams we indeed verify that the divergences
disappear by GIM as stated in section \ref{4w}. We remark that
the evanescent operators in (\ref{greek}) can in general mix into physical
operators \cite{bw}. We have checked that this is not the case here.

The results for the individual diagrams
of \fig{bi} and \ref{biqcd}
can be found in the appendix.
They represent
\begin{eqnarray}
\langle {\cal O}_{ij} \lt( \mu \r) \rangle   &=&
\langle {\cal O}_{ij} \lt( \mu \r) \rangle^{(0)}
+  \frac{\alpha (\mu) }{4 \pi }
    \langle {\cal O}_{ij} \lt( \mu \r) \rangle^{(1)}
+ O(\alpha^2) \no
\end{eqnarray}
with $\langle {\cal O}_{ij} \lt( \mu \r) \rangle^{(0)} $  given
in (\ref{tau}) and
\begin{eqnarray}
\langle {\cal O}_{ij} \lt( \mu \r) \rangle^{(1)}   &=&
                \frac{\mc (\mu)}{8 \pi^2} \lt[
                \langle \oll  \rangle ^{(0)}   a_{LL}^{(ij)} (\mu)
             +  \langle \hat{T}  \rangle ^{(0)}   \tau_{ij} h_T
              +  \langle \hat{U}  \rangle ^{(0)}   \tau_{ij} h_U \r]
              .
\label{bime}
\end{eqnarray}
Let us analyze
(\ref{bime}) first by
looking back  at the factorization at $\mu=\mu_W$:
The unphysical Dirac structures $\hat{T}$ and $\hat{U}$ come with the same
coefficient functions $h_T$ and $h_U$ defined in (\ref{ltu}) as
$\langle H ( \mu_W) \rangle ^{(1)} $ in  (\ref{h1}).
$\tau _{ij}$ has been defined in (\ref{tau}), and with
$\sum \tau_{ij} =1$ the factorization
of the
$\hat{T}$-- and $\hat{U}$--part
at $\mu=\mu_W$ becomes transparent in
(\ref{nlofacw}).
We split $a_{LL}^{(ij)}$ in (\ref{bime})
into its physical part $c_{LL}^{(ij)}$ and those parts which depend
on the infrared regulators or involve the gauge parameter:
\begin{eqnarray}
a_{LL}^{(ij)}(\mu) &=& c_{LL}^{(ij)}(\mu) + \tau _{ij} \lt\{
       \frac{N-1}{2 N} 3 \log \frac{\md \ms}{\mu ^4}  \r.     \nn
&&   \quad \quad  \quad \quad + \,
   \xi \lt[  \lt(C_F + \frac{N-1}{2 N}  \r)
    \lt( 2 - 2 \frac{\ms \log (\ms/\mu^2) -
                     \md \log (\md/\mu^2) }{\ms-\md}     \r) \r. \nn
&& \quad \quad \quad \quad \quad \quad \quad \lt. \lt. +
     \frac{N-1}{2 N} \log \frac{\md \ms}{\mu^4}  \r]   \r\} . \label{all}
\end{eqnarray}
Clearly also in (\ref{all}) the unphysical terms coincide with
those of (\ref{ltu}), so that they correctly give no contribution in
(\ref{nlofacw}) to the Wilson coefficient.

The desired physical parts in (\ref{all}) read:
\begin{eqnarray}
c_{LL}^{(++)} (\mu) &=& \tau_{++} \: 3 (1-N) \log \frac{\mc (\mu)}{\mu^2}   \nn
&&                   + \frac{102-73 N -32 N^2 +3 N^3}{8 N}  +
               \pi^2 \frac{-6+6 N+N^2 -N^3}{12 N} , \no \\[2mm]
c_{LL}^{(+-)}(\mu) \; = \; c_{LL}^{(-+)} ( \mu)
&=& \tau_{+-}  \: 3 (-1-N) \log \frac{\mc (\mu)}{\mu^2}    \nn
&&                   + \frac{34-39 N +8 N^2 -3 N^3}{8 N}  +
               \pi^2 \frac{-2+4 N-3 N^2 +N^3}{12 N} , \no  \\[2mm]
c_{LL}^{(--)}(\mu) &=& \tau_{--} \: 3 (-3-N) \log \frac{\mc
      (\mu)}{\mu^2}  \nn
&&                   + \frac{-34+19 N +12 N^2 +3 N^3}{8 N}  +
               \pi^2 \frac{2-6 N+5 N^2 -N^3}{12 N} . \label{physll}
\end{eqnarray}
With (\ref{physll}) one verifies (\ref{nlofacw}) as promised
in section \ref{4w}.

\subsubsection{Determination of the Wilson Coefficient $C(\mu)$}
Next we want to use (\ref{bime}) to
solve (\ref{4matchc}) for
$C(\mu _c)$.
To this end we also need $\langle \oll   \rangle$, which is made up
of the diagrams of \fig{loc} and \ref{locqcd}.
One has \cite{bw}
\begin{eqnarray}
\langle \oll \lt( \mu \r) \rangle   &=&
\langle \oll \lt( \mu \r) \rangle^{(0)}  + \frac{\alpha (\mu) }{4 \pi}
          \lt[
                \langle \oll  \rangle ^{(0)}   a (\mu)
             +  \langle \hat{T}  \rangle ^{(0)}    h_T
              +  \langle \hat{U}  \rangle ^{(0)}   h_U \r] . \label{lome}
\end{eqnarray}
Here $h_T$ and $h_U$ have been defined in (\ref{ltu}), and we extract
the physical part $c$ out of $a(\mu)$:
\begin{eqnarray}
a (\mu) &=& c  +  \lt\{
       \frac{N-1}{2 N} 3 \log \frac{\md \ms}{\mu ^4}  \r.     \nn
&&   \quad \quad  \quad \quad + \,
   \xi \lt[  \lt(C_F + \frac{N-1}{2 N}  \r)
    \lt( 2 - 2 \frac{\ms \log (\ms/\mu^2) -
                     \md \log (\md/\mu^2) }{\ms-\md}     \r) \r. \nn
&& \quad \quad \quad \quad \quad \quad \quad \lt. \lt. +
     \frac{N-1}{2 N} \log \frac{\md \ms}{\mu^4}  \r]   \r\} , \nn
c  &=&
 -3 \, C_F -5 \, \frac{N-1}{2 N} .  \label{acloc}
\end{eqnarray}
Now we are in a position to solve the matching condition
(\ref{4matchc}) for $C(\mu_c)$:
\begin{eqnarray}
C(\mu_c) &=&
 \! \! \mc(\mu_c ) \sum_{i,j=+,-}
           \lt( \frac{\alpha (\mu _b)}{\alpha (\mu_c )}   \r) ^{d_{ij}^{(4)}}
           \lt( \frac{\alpha (\mu _W)}{\alpha (\mu _b)}   \r) ^{d_{ij}^{(5)}}
\lt\{ \tau_{ij}
\lt[ 1
   + \frac{\alpha (\mu _b) - \alpha (\mu_c )}{4 \pi } Z_{ij}^{(4)} \r. \r. \nn
&&   \quad \quad \lt. \lt.
  + \frac{\alpha (\mu _W)- \alpha (\mu _b)}{4 \pi } Z_{ij}^{(5)}
  + \frac{\alpha (\mu _W) }{4 \pi } b_{ij} (\mu_W)
        \r]
  + \frac{\alpha(\mu_c)}{4 \pi}
       r_{ij} (\mu_c) \r\} . \label{wccloc}
\end{eqnarray}
Here
\begin{eqnarray}
b_{ij} (\mu_W) &=&
   \frac{\gamma _{ij}^{(0)}}{2} \log \frac{\mu _W^2}{\mw} + B_i + B_j
 ,\quad i,j=+,-,    \label{4bij}
\end{eqnarray}
where $B_\pm$ defined in (\ref{bpm}) originates from
the matching at $\mu_W$,
while $\tau_{ij}$ and
\begin{eqnarray}
r_{ij} (\mu_c) &=& c_{LL}^{ij} (\mu_c)- \tau_{ij} c  \label{rij}
\end{eqnarray}
have entered the result at the c--quark threshold $\mu_c$.
All other quantities in (\ref{wccloc}) are related to the
RG evolution.

Notice that again those parts of $\langle {\cal O}_{ij} \rangle$
and $\langle \oll \rangle$ which depend on the IR structure
or the gauge parameter have not contributed to the Wilson
coefficient.
By working in an arbitrary $R_\xi$--gauge for the gluon we have
explicitly proven that our final Wilson coefficient
$C(\mu)$ is gauge
independent.
The use of small quark masses $m_s$ and $m_d$
as infrared regulators
has enabled us to verify that $C(\mu)$ does not depend
on the infrared structure.
This would not have been so,
if we had  regulated
both UV and IR
singularities dimensionally, because
the infrared structure would  not have been explicit and
the factorizations
would have been trivial due to the absence of the
operators  $\hat{T}$ and $\hat{U}$.

$C(\mu)$ for $\mu < \mu_c$ is readily obtained from (\ref{belowc}).
The relation of $C(\mu)$  to $\eta_1$ is contained in
(\ref{s2}), (\ref{hmc}) and (\ref{smallb}):
\begin{eqnarray}
 C(\mu)&=& \eta_1 (\mu_W, \mu_b, \mu_c) \mc (\mu_c) b (\mu) , \no
\end{eqnarray}
so that we can  write down the final result now.

\subsubsection{The Final Result for $\eta_1$ in Next--To--Leading Order}
\label{nlofinalresult}
The coefficient $\eta_1$ in (\ref{s2}) reads in next--to--leading order:
\begin{eqnarray}
\eta_1 \! \! &=&
 \! \! \lt( \alpha (\mu_c) \r)^{d_+^{(3)}}
   \! \! \! \sum_{i,j=+,-} \!
           \lt( \frac{\alpha (\mu _b)}{\alpha (\mu_c )}   \r) ^{d_{ij}^{(4)}}
  \! \! \!   \lt( \frac{\alpha (\mu _W)}{\alpha (\mu _b)}   \r) ^{d_{ij}^{(5)}}
 \! \lt\{ \tau_{ij} \!
\lt[ 1
   + \frac{\alpha (\mu _c ) }{4 \pi } Z_+^{(3)}
   + \frac{\alpha (\mu _b) - \alpha (\mu_c )}{4 \pi } Z_{ij}^{(4)} \r.
  \r. \no \\[2mm]
&& \quad \quad \quad \quad \quad \lt. \lt.
  + \frac{\alpha (\mu _W)- \alpha (\mu _b)}{4 \pi } Z_{ij}^{(5)}
  + \frac{\alpha (\mu _W) }{4 \pi } b_{ij} (\mu_W)
        \r] + \frac{\alpha(\mu_c)}{4 \pi} r_{ij} (\mu_c) \r\} .
\label{4eta1}
\end{eqnarray}

The following table sums up in which equations the various quantities in
(\ref{4eta1}) are defined:
\begin{displaymath}
\begin{array}{|c|c|c|c|c|c|c|c|}
\hline
&&&&&&&\\[-4mm]
\alpha (\mu ) & d_+^{(3)} & d_{ij}^{(f)} & \tau_{ij} & Z_+^{(3)} &
Z_{ij}^{(f)} & b_{ij} & r_{ij} \\[2mm]\hline
&&&&&&&\\[-4mm]
(\ref{runalpha}) & (\ref{dz}) & (\ref{dz}) & (\ref{tau}) & (\ref{dz})
& (\ref{dz}) & (\ref{4bij}) & (\ref{rij}), (\ref{physll}),(\ref{acloc})
\\[2mm]\hline
\end{array}
\end{displaymath}

Let us come back to  the scale  ambiguity at $\mu_c$,
which we had also discussed at the LL result
by expanding $\eta_1 (\mu_c) \mc (\mu_c)$
around $\mu_c =m_c^* = m_c (m_c)$
to order $\alpha \log \lt( \mu_c^2 / ( m_c^* )^2  \r) $
in (\ref{explog}).
In the NLO  expression (\ref{4eta1}) we encounter
via $r_{ij}$ additional contributions to  the coefficient
of  $\alpha \log \lt( \mu_c^2 / ( m_c^* )^2  \r) $.
They enter $r_{ij}$ in (\ref{rij}) via $c_{LL}^{(ij)}$ given
in (\ref{physll}).
Consider for example the coefficient of the logarithm in
$c_{LL}^{(--)}$. It reads
\begin{eqnarray}
\tau_{--}\: 3\: (-3-N) &=& \tau_{--}
	\lt( \frac{\gamma_{--}^{(0)} -\gamma_+^{(0)}}{2}
               - \gamma_m ^{(0)}    \r)  \no
\end{eqnarray}
and thereby cancels the corresponding coefficient in (\ref{explog}).
The same holds for the other terms in the sum so that in the NLO result
(\ref{4eta1}) the scale  ambiguity at $\mu_c$ is removed
to order $\alpha$.

We have expanded (\ref{4eta1}) further to estimate the remaining
dependence  on $\mu_c$. At order
$\alpha^2 \log \lt( \mu_c^2 / ( m_c^* )^2  \r) $
we find a large contribution of the $(--)$--term,
which originates in the fact that the anomalous dimension
of ${\cal O}_{--}$ is negative and large in magnitude.
In section \ref{5} we will discuss the scale
ambiguities numerically.

We close this section by mentioning another check of our result:
If one switches off  the running between
$\mu_W$ and $\mu_c$ in (\ref{4eta1}) by expanding
$[\alpha(\mu_W)/\alpha (\mu_b)]$ etc.\ around $\alpha(\mu_W)$,
one must get the same result as by expanding $\eta_2$
calculated in \cite{bjw} in $m_t^2/\mw $ and setting $m_t=m_c$.
This is indeed the case.

%
%
\section{Numerical Results}
\label{5}

In this section we will point out the numerical impact of our NLO calculation
on the value of $\eta_1$ and the $K_L$ -- $K_S$ mass difference.
Special attention is paid to the dependence on the matching scales $\mu_c$
and $\mu_W$.

The charm quark contribution to the $\Delta S = 2$ Hamiltonian was given in
(\ref{hcc}).
Replacing the $S$ function therein by its lowest nonvanishing order in the
expansion with respect to $\mc / \mw$ leads to
\begin{eqnarray}
H^c (\mu) &=&  \frac{G_F^2}{16 \pi^2} \lambda_c^2 \cdot
                 \eta_1 (\mu_W,\mu _b, \mu_c)
         \cdot     \mc(\mu_c) \cdot b(\mu) \cdot \oll(\mu )
\nn
          &=&  \frac{G_F^2}{16 \pi^2} \lambda_c^2 \cdot
                 \eta_1^{\star} (\mu_W,\mu _b, \mu_c)
         \cdot     m_c^{\star 2} \cdot b(\mu) \cdot \oll(\mu )
\label{hccnum}
\end{eqnarray}
with
\begin{eqnarray}
\eta_1^{\star}(\mu_W, \mu_b, \mu_c) &=&
\frac{\mc(\mu_c)}{m_c^{\star 2}} \;
\eta_1(\mu_W, \mu_b, \mu_c),
\nn
m_c^{\star} &=&
m_c(m_c).
\end{eqnarray}
$m_c(\mu_c)$ needed in (\ref{hccnum}) can be obtained from $m_c^{\star}$
by using the mass evolution equation (\ref{rum}).

$H^c(\mu)$ is directly related to physical observables and therefore
must not depend on the scales $\mu_W$, $\mu_b$, $\mu_c$.
For this reason we have introduced $\eta_1^{\star}$ which multiplies
only quantities which do not depend on these scales.
$\eta_1^{\star}$ is only relevant for the discussion of the dependence
on $\mu_c$, since for $\mu_c = m_c^{\star}$ one has
$\eta_1 \equiv \eta_1^{\star}$.

Let us recall the expression for $\eta_1$
\begin{eqnarray}
\eta_1 \! \! &=&
 \! \! \lt( \alpha (\mu_c) \r)^{d_+^{(3)}}
   \! \! \! \sum_{i,j=+,-} \!
           \lt( \frac{\alpha (\mu _b)}{\alpha (\mu_c )}   \r) ^{d_{ij}^{(4)}}
  \! \! \!   \lt( \frac{\alpha (\mu _W)}{\alpha (\mu _b)}   \r) ^{d_{ij}^{(5)}}
 \! \lt\{ \tau_{ij} \!
\lt[ 1
   + \frac{\alpha (\mu _c ) }{4 \pi } Z_+^{(3)}
   + \frac{\alpha (\mu _b) - \alpha (\mu_c )}{4 \pi } Z_{ij}^{(4)} \r. \r. \nn
&&   \quad \quad \quad \quad \quad \quad \lt. \lt.
  + \frac{\alpha (\mu _W)- \alpha (\mu _b)}{4 \pi } Z_{ij}^{(5)}
  + \frac{\alpha (\mu _W) }{4 \pi } b_{ij} (\mu_W)
        \r] + \frac{\alpha(\mu_c)}{4 \pi} r_{ij} (\mu_c) \r\}
\label{numreseta1}
\end{eqnarray}
and list its ingredients for the case $N = 3$:
\begin{displaymath}
d^{(3)}_{+} = \frac{2}{9}, \quad
Z^{(3)}_{+} = - \frac{307}{162},
\end{displaymath}
\begin{displaymath}
\begin{array}{||c||c|c|c||}
\hline
\hline
ij & ++ & +-\; =\; -+ & -- \\
\hline
\hline
\ & \ & \ & \ \\[-4mm]
d^{(4)}_{ij} & \frac{12}{25} & - \frac{6}{25} & - \frac{24}{25} \\[1mm]
d^{(5)}_{ij} & \frac{12}{23} & - \frac{6}{23} & - \frac{24}{23} \\[1mm]
\hline
\ & \ & \ & \ \\[-4mm]
Z^{(4)}_{ij} &
	- \frac{6719}{1875} & \frac{419}{3750} & \frac{7138}{1875} \\[1mm]
Z^{(5)}_{ij} &
	- \frac{5165}{1587} & - \frac{631}{3174} & \frac{4534}{1587} \\[1mm]
\hline
\ & \ & \ & \ \\[-4mm]
\tau_{ij} & \frac{3}{2} & -\frac{1}{2} & \frac{1}{2} \\[1mm]
\hline
\ & \ & \ & \ \\[-4mm]
r_{ij}(\mu_c) &
	-9 \log \frac{\mc}{\mu_c^2} - 5 - \frac{\pi^2}{6} &
	6 \log \frac{\mc}{\mu_c^2} - \frac{20}{3} + \pi^2 \frac{5}{18} &
	-9 \log \frac{\mc}{\mu_c^2} + \frac{35}{3} + \frac{\pi^2}{18} \\[2mm]
\hline
\ & \ & \ & \ \\[-4mm]
b_{ij}(\mu_W) &
	4 \log \frac{\mu_W^2}{\mw} + \frac{22}{3} &
	-2 \log \frac{\mu_W^2}{\mw} - \frac{11}{3} &
	-8 \log \frac{\mu_W^2}{\mw} - \frac{44}{3} \\[2mm]
\hline
\hline
\end{array}
\end{displaymath}

\subsection{Numerical Results for $\eta_1$ and $\Delta m_K$}
\label{numres}
This section lists the numerical values for $\eta_1$ and the
$K_L$ -- $K_S$ mass difference $\Delta m_K$.
The scale ambiguities will be viewed at in sect.\ \ref{numscaledep}.

For our numerical estimates
we will use $\laMSb = \Lambda_4$
and the following list of input parameters \cite{pdg}:
\begin{eqnarray}
 & &
M_W = 80 \gev, \quad \mu_b = m_b = 4.8 \gev,
\nn
 & &
B_K = 0.7, \quad m_K = 0.498 \gev, \quad f_K = 0.161 \gev, \quad
(\Delta m)_K^{\rm EXP} = 3.522 \cdot 10^{-15} \gev,
\nn
 & &
{\rm Re} \lambda_c = 0.215, \quad \gf = 1.16639 \cdot 10^{-5} \gev^{-2}.
\label{numdefault}
\end{eqnarray}

In this section we will set the matching scales
$\mu_c = m_c^{\star}$ and $\mu_W = M_W$.
Table \ref{eta1lamc} shows the LO and NLO result for $\eta_1$ for various
values of $m_c$ and $\laMSb$.

\begin{figure}[htb]
\centerline{
\rotate[r]{
\epsfysize=14cm
\epsffile{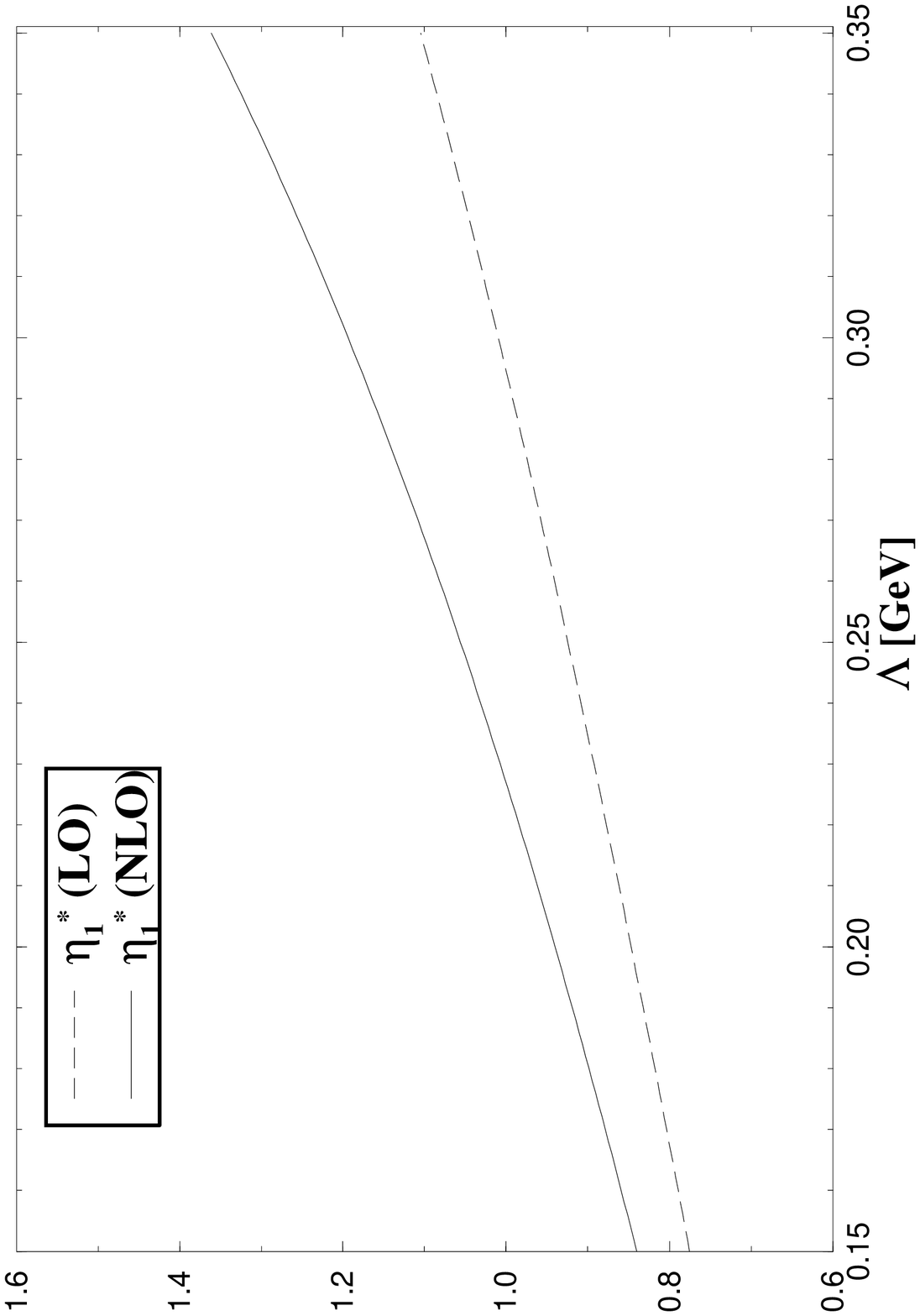}
}}
\caption[]{The dependence of $\eta_1^{\star}$ on $\Lambda_4 = \laMSb$.
}
\label{figeta1laMSb}
\end{figure}

\begin{table}[htb]
%
%
\begin{tabular}{||c||c|c||c|c||c|c||c|c||c|c||}
\hline
\hline
 & \multicolumn{2}{c||}{\ }
 & \multicolumn{2}{c||}{\ }
 & \multicolumn{2}{c||}{\ }
 & \multicolumn{2}{c||}{\ }
 & \multicolumn{2}{c||}{\ }
\\[-4mm]
$\Lambda_{\overline {\rm MS}}$
 & \multicolumn{2}{c||}{0.150}
 & \multicolumn{2}{c||}{0.200}
 & \multicolumn{2}{c||}{0.250}
 & \multicolumn{2}{c||}{0.300}
 & \multicolumn{2}{c||}{0.350}
\\[1mm]
\hline
\hline
$m_c^{\star}$ & LO & NLO & LO & NLO & LO & NLO & LO & NLO & LO & NLO\\
\hline
\hline
1.25 &  0.809 &  0.885 &  0.895 &  1.007 &  0.989 &  1.154 &  1.096 &  1.334 &
1.216 &  1.562 \\
1.30 &  0.797 &  0.868 &  0.877 &  0.982 &  0.965 &  1.117 &  1.064 &  1.281 &
1.175 &  1.485 \\
\hline
1.35 &  0.786 &  0.854 &  0.861 &  0.960 &  0.944 &  1.085 &  1.035 &  1.235 &
1.138 &  1.419 \\
1.40 &  0.775 &  0.840 &  0.847 &  0.940 &  0.924 &  1.056 &  1.010 &  1.194 &
1.105 &  1.361 \\
1.45 &  0.766 &  0.828 &  0.834 &  0.922 &  0.907 &  1.030 &  0.987 &  1.157 &
1.075 &  1.310 \\
\hline
1.50 &  0.757 &  0.817 &  0.822 &  0.905 &  0.890 &  1.006 &  0.966 &  1.125 &
1.048 &  1.265 \\
1.55 &  0.749 &  0.806 &  0.810 &  0.890 &  0.876 &  0.985 &  0.946 &  1.095 &
1.024 &  1.225 \\
\hline
\hline
\end{tabular}
\caption[]{
	$\eta_1$ to LO and NLO for different values of
	$\Lambda_4 = \laMSb$ and
	$\mu_c = m_c^{\star}$ (both given in $\gev$).
}
\label{eta1lamc}
\end{table}
{}From this table we see, that the NLO correction is positive and
sizable.
For a typical value of $\laMSb = 0.3 \gev$ it amounts to about
$20$ percent of the LO result.
The dependence of $\eta_1$ on $\laMSb$ turns out to be stronger when
NLO effects are included.
This can be seen from \fig{figeta1laMSb}, where we have taken
$m_c^{\star} = 1.4 \gev$.
Larger values of $\laMSb$ result in larger corrections and vice versa.
The absolute value of $\eta_1$ decreases with increasing $m_c$.

Next, using (\ref{numdefault}), we calculate the contribution of $H^c$ to the
$K_L$ -- $K_S$ mass difference \cite{bss}
\begin{eqnarray}
(\Delta m)_K^{c} &=&
\frac{\gf^2}{6 \pi^2}\: m_K \: f_K^2\: B_K\: ({\rm Re} \lambda_c)^2 \:
	m_c^{\star 2}\: \eta_1^{\star}
\end{eqnarray}
with the result listed in table \ref{dmlamc}.

\addtolength{\tabcolsep}{-1mm}
\begin{table}[htb]
%
%
%
\begin{tabular}{||c||c|c||c|c||c|c||c|c||c|c||}
\hline
\hline
 & \multicolumn{2}{c||}{\ }
 & \multicolumn{2}{c||}{\ }
 & \multicolumn{2}{c||}{\ }
 & \multicolumn{2}{c||}{\ }
 & \multicolumn{2}{c||}{\ }
\\[-4mm]
$\Lambda_{\overline {\rm MS}}$
 & \multicolumn{2}{c||}{0.150}
 & \multicolumn{2}{c||}{0.200}
 & \multicolumn{2}{c||}{0.250}
 & \multicolumn{2}{c||}{0.300}
 & \multicolumn{2}{c||}{0.350}
\\[1mm]
\hline
\hline
 & & & & & & & & & &\\[-4mm]$m_c^{\star}$ & $\Delta m$ & $\frac{\Delta
m}{\Delta m_{\rm EXP}}$ & $\Delta m$ & $\frac{\Delta m}{\Delta m_{\rm EXP}}$ &
$\Delta m$ & $\frac{\Delta m}{\Delta m_{\rm EXP}}$ & $\Delta m$ & $\frac{\Delta
m}{\Delta m_{\rm EXP}}$ & $\Delta m$ & $\frac{\Delta m}{\Delta m_{\rm
EXP}}$\\[1mm]
\hline
\hline
1.25 &  1.327 &  0.377 &  1.510 &  0.429 &  1.730 &  0.491 &  2.000 &  0.568 &
2.342 &  0.665 \\
1.30 &  1.408 &  0.400 &  1.593 &  0.452 &  1.812 &  0.514 &  2.078 &  0.590 &
2.409 &  0.684 \\
\hline
1.35 &  1.493 &  0.424 &  1.679 &  0.477 &  1.897 &  0.539 &  2.159 &  0.613 &
2.481 &  0.705 \\
1.40 &  1.580 &  0.449 &  1.768 &  0.502 &  1.986 &  0.564 &  2.245 &  0.637 &
2.560 &  0.727 \\
1.45 &  1.670 &  0.474 &  1.860 &  0.528 &  2.078 &  0.590 &  2.335 &  0.663 &
2.643 &  0.751 \\
\hline
1.50 &  1.763 &  0.501 &  1.955 &  0.555 &  2.173 &  0.617 &  2.428 &  0.689 &
2.731 &  0.776 \\
1.55 &  1.859 &  0.528 &  2.052 &  0.583 &  2.271 &  0.645 &  2.525 &  0.717 &
2.824 &  0.802 \\
\hline
\hline
\end{tabular}
\caption[]{
	$(\Delta m)_K^c$ to NLO in units of $10^{-15} \gev$
	for different values of $\Lambda_4 = \laMSb$ and
	$\mu_c = m_c^{\star}$ (both given in $\gev$).
	$B_K = 0.7$.
}
\label{dmlamc}
\end{table}
\addtolength{\tabcolsep}{1mm}
Since $(\Delta m)_K^{c}$ depends linearly on the $B_K$,
the result can easily be rescaled to other values of this parameter.

Note that for $m_c^{\star} = 1.4 \gev$ and $\laMSb = 0.3 \gev$
the contribution of $H^c$ to $(\Delta m)_K^{\rm EXP}$ is about 64 percent.
If we also take into account the top quark contribution in
$H^{\Delta S=2}$ (\ref{s2}), which is about 6 percent,
the complete short distance analysis reproduces about 70 percent of
$(\Delta m)_K^{\rm EXP}$.

Using vacuum insertion $B_K = 1$,
the charm and top quark short distance contribution reproduces the full
experimental mass difference,
but \cite{gks,bbg,d,pr} favour a lower value for $B_K$.

\subsection{Scale Dependences of $\eta_1$}
\label{numscaledep}

\begin{figure}[htb]
\centerline{
\rotate[r]{
\epsfysize=14cm
\epsffile{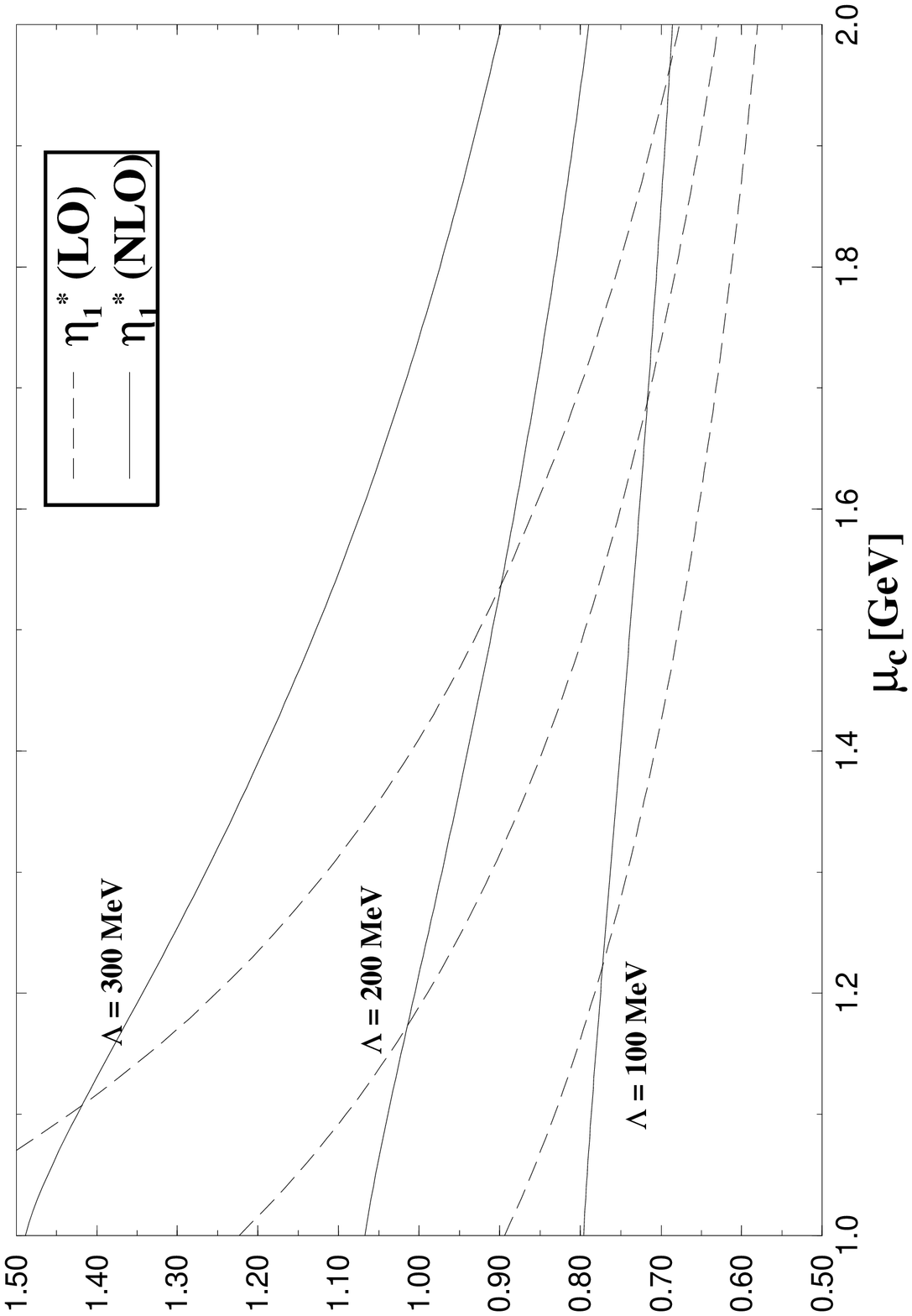}
}}
\caption[]{The dependence of $\eta_1^{\star}$
on the matching scale $\mu_c$, where the $c$ quark is removed from the theory
for $\laMSb = 0.1,\, 0.2,\, 0.3 \gev$.
$m_c^{\star} = 1.4 \gev$ has been kept fixed.
The other scales have been chosen to be
$\mu_b = 4.8 \gev$ and $\mu_W = M_W = 80 \gev$ .}
\label{figeta1muc}
\end{figure}

Let us now discuss the dependence of $\eta_1^{\star}$ on the matching
scale $\mu_c$.
We will fix the values of
$m_c(m_c) = m_c^{\star} = 1.4 \gev$,
$\mu_b = 4.8 \gev$
and $\mu_W = M_W = 80 \gev$.
As can be seen from \fig{figeta1muc}
the scale dependence in the NLO calculation is reduced quite considerably
compared to the LO result.

If we take as a reference value $\laMSb = 0.3 \gev$, then
varying $\mu_c$ from $1.1$ to $1.7 \gev$ amounts to a change of $63 \%$
in the LO case and $34 \%$ in the NLO calculation:
\begin{displaymath}
\eta_1^{\star} \; = \; 1.01
\begin{array}{l}
\footnotesize +0.42 \\[-.7mm]
\footnotesize -0.21
\end{array}
\quad \mbox{in LO }, \quad
\eta_1^{\star} \;=\; 1.19
\begin{array}{l}
\footnotesize +0.23 \\[-.7mm]
\footnotesize -0.17
\end{array}
\quad \mbox{in NLO }. \no
\end{displaymath}
For $\laMSb = 0.2 \gev$ the corresponding variations are
$45\%$ in LO and $20\%$ in NLO:
\begin{eqnarray}
\eta_1^{\star} &=& 0.85
\begin{array}{l}
\footnotesize +0.24 \\[-.7mm]
\footnotesize -0.13
\end{array}
\quad \mbox{in LO }, \quad
\eta_1^{\star} \; = \; 0.94
\begin{array}{l}
\footnotesize +0.10 \\[-.7mm]
\footnotesize -0.08
\end{array}
\quad \mbox{in NLO }. \no
\end{eqnarray}
The variation of $\mu_c$ between $1.1$ and $1.7 \gev$ is quite
conservative and may overestimate the `theoretical error'.
Note, that setting $\mu_c \approx 1.25 \gev$ in the LO expression already
reproduces the NLO result for $\mu_c = 1.4 \gev$ for all reasonable values
of $\laMSb$.

We will further analyse the remaining scale setting ambiguity by looking
at the individual contributions coming from the different terms in the
sum in (\ref{numreseta1})
as we already did analytically in sec.\ \ref{nlofinalresult}.

\begin{figure}[htb]
\centerline{
\rotate[r]{
\epsfysize=14cm
\epsffile{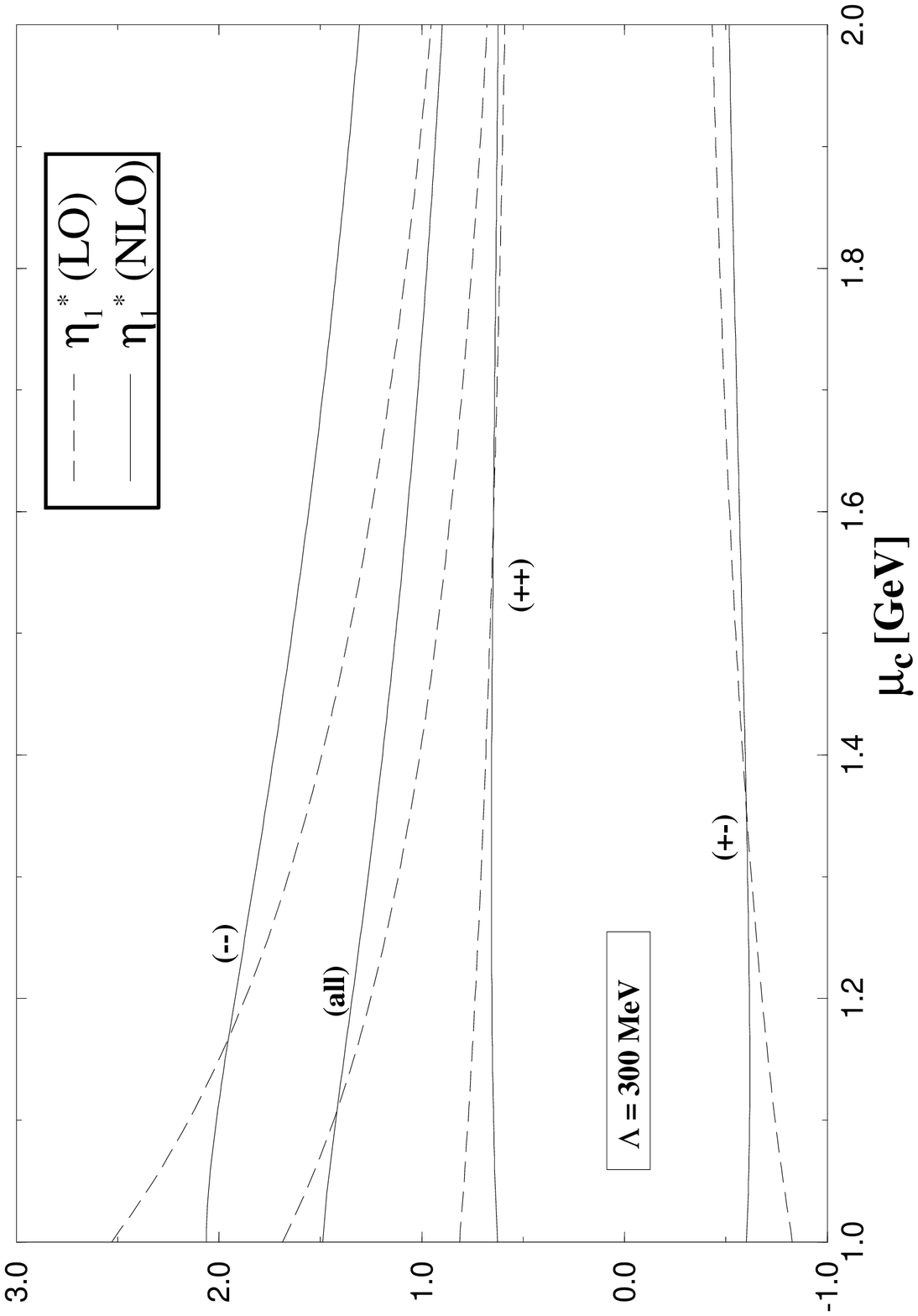}
}}
\caption[]{The dependence of the $(++)$, $(+-)$ and $(--)$ contributions
to $\eta_1^{\star}$ on $\mu_c$ for $\laMSb = 0.3 \gev$.
The line labeled by ``all'' is obtained by adding $(++) + 2 (+-) + (--)$.
}
\label{figeta1indivmuc}
\end{figure}
In \fig{figeta1indivmuc} one can easily see, that the $(++)$ and
$(+-)$ contributions behave essentially flat as they should, but
they contribute with opposite sign to $\eta_1^{\star}$ and therefore
almost cancel in the sum.
On the other hand the $(--)$ part carries a large residual scale
dependence and is the largest of the three contributions.

Let us now continue with the discussion of residual scale dependences in
$\eta_1^{\star}$ by varying the scale $\mu_W$.
We will fix $\mu_c = m_c^{\star} = 1.4 \gev$,
$\mu_b = 4.8 \gev$, $M_W = 80 \gev$.
Also here --- as in the case of $\mu_c$ discussed above ---
the scale dependence is reduced considerably by the inclusion of NLO
effects.
Taking $\laMSb = 0.3 \gev$ and varying $\mu_W$ from $60$ to $100 \gev$
we find a change of $\eta_1^{\star}$ by $7 \%$ in LO and $4 \%$ in NLO.
This residual scale dependence therefore turns out to be small compared
to the one found for $\mu_c$.

$\eta_1^{\star}$ turns out to be very stable when varying the $b$--scale
$\mu_b$.
Even if we neglect completely the effects from an effective 5 quark
theory by setting $\mu_b = \mu_W$,
therefore evolving from scale $\mu_W$ down to $\mu_c$
in an effective 4 quark theory, the error turns out to be of
the order of 1 percent.
This provides us with a simple and useful approximation for the numerical
evaluation of $\eta_1$.
\begin{eqnarray}
\eta_1 \! \! & \approx &
 \! \! \lt( \alpha (\mu_c) \r)^{d_+^{(3)}}
   \! \! \! \sum_{i,j=+,-} \!
           \lt( \frac{\alpha (\mu _W)}{\alpha (\mu_c )}   \r) ^{d_{ij}^{(4)}}
 \! \lt\{ \tau_{ij} \!
\lt[ 1
   + \frac{\alpha (\mu _c ) }{4 \pi } Z_+^{(3)}
   + \frac{\alpha (\mu _W) - \alpha (\mu_c )}{4 \pi } Z_{ij}^{(4)} \r. \r. \nn
&&   \quad \quad \quad \quad \quad \quad \quad \quad \quad \quad
     \quad \quad \lt. \lt.
  + \frac{\alpha (\mu _W) }{4 \pi } b_{ij} (\mu_W)
        \r] + \frac{\alpha(\mu_c)}{4 \pi} r_{ij} (\mu_c) \r\}
\label{numapproxeta1}
\end{eqnarray}


%
%
\section{Conclusions}
In the present work we have calculated the complete
next--to--leading order short distance QCD corrections to
the coefficient $\eta_1$ of the
effective $\Delta S=2$ hamiltonian describing the
$K^0$--$\ov{K^0}$ mixing.
We have used  dimensional
regularization and the  $\ov{{\rm MS}}$--scheme with
anticommuting $\gamma_5$.
Large logarithms have been summed by factorizing the
Feynman amplitudes and applying the renormalization group
evolution to the Wilson coefficients.
To this end the renormalization  of bilocal
operators had to be  studied in detail.
We have explicitly proven that all Wilson coefficients
involved in the calculation depend neither on the gluon gauge
nor on the infrared regulators.
We have further checked that the dependence of the
leading order result
on the renormalization scales
is removed at NLO to order $\alpha$.

The NLO short distance QCD corrections enhance $\eta_1$ by
20 \% with respect to the LO result.
The short distance part of the
$K_L$--$K_S$ mass difference is increased by the same amount,
because it receives its dominant contribution
from the first two quark generations.
 Finally the theoretical uncertainty  due to the remaining
dependence on the
renormalization scale has been reduced by a factor of
2 compared to  the LO result.

\subsection*{Acknowledgements}
At first we would like to thank A.\ J.\  Buras for
suggesting this project and for
his permanent
encouragement during our work.
We are grateful to him and G.\ Buchalla for carefully reading the
manuscript and appreciate
many helpful discussions with them,
M.\ E.\ Lautenbacher and M.\ Misiak.

%
%
\appendix
\section{The Result for the Individual Diagrams}

In this appendix we collect the results for the individual diagrams depicted
in fig.\ \ref{bi} and fig.\ \ref{biqcd} including their QCD counterterms.
The GIM mechanism has already been incorporated in the form
$(c,c) - (u,c) - (c,u) + (u,u)$,
where $(k,l)$ denotes a diagram with internal $k$ and $l$ quarks.
Then we get for the insertion of the bilocal operator ${\cal O}^{ij}$
\begin{eqnarray}
{\rm D_0}^{ij} & = &
	-i \frac{1}{16 \pi^2} m_c^2 \; {\rm D_0} \;
	\left({\cal C}_0^{ij} \; \1 + {\tilde {\cal C}_0^{ij}} \; \tw \right),
	\nn
{\rm D_\ell}^{ij} & = &
	-i \frac{1}{16 \pi^2} m_c^2 \frac{\alpha}{4 \pi}
	\left({\rm D_\ell^{\xi =0}} + \xi \;
		{\rm D_\ell^{gauge}} \right) \;
	\left({\cal C}_{\ell}^{ij} \; \1 +
		{\tilde {\cal C}_{\ell}^{ij}} \; \tw \right),
	\quad \ell=1,\ldots,7,
	\nn
{\rm D_8}^{ij} & = & 0,
\label{diagindiveff}
\end{eqnarray}
where ${\cal C}_\ell^{ij}$ and ${\tilde {\cal C}_\ell^{ij}}$
denote colour factors listed in table\ \ref{ColorTable}.
The gauge independent part
${\rm D}_\ell^{\xi=0}$
and the one proportional to the gluon gauge parameter
${\rm D}_\ell^{\rm gauge}$
read
\begin{eqnarray}
{\rm D_{0}} & = &
+\gamma_{\mu}L \otimes \gamma^{\mu}L\left[
1
\right] \nonumber \\
 & &
\nonumber \\
{\rm D_{1}^{\xi=0}} & = &
+\gamma_{\mu}R \otimes \gamma^{\mu}L\left[
{{3\,{\rm R_1}(m_s^2,m_d^2)\,\frac{m_s m_d}{\mu^2}}\over 2}
\right] \nonumber \\
 & &
+\gamma_{\mu}L \otimes \gamma^{\mu}L\left[
{3\over 2}
\right] \nonumber \\
 & &
\nonumber \\
{\rm D_{1}^{gauge}} & = &
+\gamma_{\mu}R \otimes \gamma^{\mu}L\left[
{{{\rm R_1}(m_s^2,m_d^2)\,\frac{m_s m_d}{\mu^2}}\over 2}
\right] \nonumber \\
 & &
+\gamma_{\mu}L \otimes \gamma^{\mu}L\left[
3 - {{{{\pi }^2}}\over 3} - {\rm S_1}(m_s^2,m_d^2) +
  \log (\frac{m_c^2}{\mu^2})
\right] \nonumber \\
 & &
\nonumber \\
{\rm D_{2}^{\xi=0}} & = &
+R \otimes L\left[
-3\,{\rm R_1}(m_s^2,m_d^2)\,\frac{m_s m_d}{\mu^2}
\right] \nonumber \\
 & &
+\gamma_{\mu}L \otimes \gamma^{\mu}L\left[
-{3\over 2}
\right] \nonumber \\
 & &
\nonumber \\
{\rm D_{2}^{gauge}} & = &
+R \otimes L\left[
-\left( {\rm R_1}(m_s^2,m_d^2)\,\frac{m_s m_d}{\mu^2} \right)
\right] \nonumber \\
 & &
+\gamma_{\mu}L \otimes \gamma^{\mu}L\left[
1 - {\rm S_1}(m_s^2,m_d^2)
\right] \nonumber \\
 & &
\nonumber \\
{\rm D_{3}^{\xi=0}} & = &
+R \otimes R\left[
-3
\right] \nonumber \\
 & &
+\gamma_{\mu}L \otimes \gamma^{\mu}L\left[
-6 + 3\,\log(\frac{m_s^2}{\mu^2}) + {{\pi }^2} - 3\,\log (\frac{m_c^2}{\mu^2})
\right] \nonumber \\
 & &
+\sigma_{\mu\nu} \otimes \sigma^{\mu\nu}\left[
{3\over 2}
\right] \nonumber \\
 & &
+\epsilon_{\mu\nu\eta\varphi} \sigma^{\mu\nu} \otimes
  \sigma^{\eta\varphi}\left[
{{-3\,i}\over 4}
\right] \nonumber \\
 & &
\nonumber \\
{\rm D_{3}^{gauge}} & = &
+R \otimes R\left[
-1
\right] \nonumber \\
 & &
+\gamma_{\mu}L \otimes \gamma^{\mu}L\left[
-2 + \log(\frac{m_s^2}{\mu^2}) + {{{{\pi }^2}}\over 3} -
  \log (\frac{m_c^2}{\mu^2})
\right] \nonumber \\
 & &
+\sigma_{\mu\nu} \otimes \sigma^{\mu\nu}\left[
{1\over 2}
\right] \nonumber \\
 & &
+\epsilon_{\mu\nu\eta\varphi} \sigma^{\mu\nu} \otimes
  \sigma^{\eta\varphi}\left[
{{-i}\over 4}
\right] \nonumber \\
 & &
\nonumber \\
{\rm D_{4}^{\xi=0}} & = &
+\gamma_{\mu}L \otimes \gamma^{\mu}L\left[
10 - {{2\,{{\pi }^2}}\over 3}
\right] \nonumber \\
 & &
\nonumber \\
{\rm D_{4}^{gauge}} & = &
+\gamma_{\mu}L \otimes \gamma^{\mu}L\left[
-4 + {{2\,{{\pi }^2}}\over 3} - 2\,\log (\frac{m_c^2}{\mu^2})
\right] \nonumber \\
 & &
\nonumber \\
{\rm D_{5}^{\xi=0}} & = &
+\gamma_{\mu}L \otimes \gamma^{\mu}L\left[
-2 - 3\,\log (\frac{m_c^2}{\mu^2})
\right] \nonumber \\
 & &
\nonumber \\
{\rm D_{5}^{gauge}} & = &
+\gamma_{\mu}L \otimes \gamma^{\mu}L\left[
2 - {{{{\pi }^2}}\over 3} + \log (\frac{m_c^2}{\mu^2})
\right] \nonumber \\
 & &
\nonumber \\
{\rm D_{6}^{\xi=0}} & = &
0
\nonumber \\
{\rm D_{6}^{gauge}} & = &
+\gamma_{\mu}L \otimes \gamma^{\mu}L\left[
-2 + {{{{\pi }^2}}\over 3} - \log (\frac{m_c^2}{\mu^2})
\right] \nonumber \\
 & &
\nonumber \\
{\rm D_{7}^{\xi=0}} & = &
+\gamma_{\mu}L \otimes \gamma^{\mu}L\left[
-7 + 3\,\log (\frac{m_c^2}{\mu^2})
\right] \nonumber \\
 & &
\nonumber \\
{\rm D_{7}^{gauge}} & = &
+\gamma_{\mu}L \otimes \gamma^{\mu}L\left[
2 - {{{{\pi }^2}}\over 3} + \log (\frac{m_c^2}{\mu^2})
\right] \nonumber \\
 & &
\nonumber \\
\end{eqnarray}
where the external spinors have been omitted.
The mass singularities are contained in
\begin{eqnarray}
{\rm R}_1 \left(m_s^2,m_d^2\right)
 & = &
\mu^2
\frac{ \log\left(\frac{m_s^2}{\mu^2}\right) -
	\log\left(\frac{m_d^2}{\mu^2}\right)}{
	m_s^2-m_d^2 },
\nn
{\rm S}_1 \left(m_s^2,m_d^2\right)
 & = &
\frac{ m_s^2 \log\left(\frac{m_s^2}{\mu^2}\right) -
	m_d^2 \log\left(\frac{m_d^2}{\mu^2}\right)}{
	m_s^2-m_d^2 }.
\end{eqnarray}
The expressions of diagrams generated from the generic ones in
fig.\ \ref{biqcd}
through a rotation by 180 degrees or a left--right--reflection
are obtained from eq.\ (\ref{diagindiveff})
according to the following rules:
\begin{itemize}
\item [${\rm D}_1^{ij}$:]
	$ \gamma_{\mu} R \otimes \gamma^{\mu} L
	\quad \longrightarrow \quad
	\gamma_{\mu} L \otimes \gamma^{\mu} R $,
\item [${\rm D}_2^{ij}$:]
	$ R \otimes L
	\quad \longrightarrow \quad
	L \otimes R $,
\item [${\rm D}_3^{ij}$:]
	$m_s \longrightarrow m_d$, \\
	$ R \otimes R
	\quad \longrightarrow \quad
	L \otimes L $, \\
	$ \varepsilon_{\mu\nu\eta\varphi}
	\sigma^{\mu\nu} \sigma^{\eta\varphi}
	\quad \longrightarrow \quad
	- \varepsilon_{\mu\nu\eta\varphi}
	\sigma^{\mu\nu} \sigma^{\eta\varphi} $.
\end{itemize}
To get the expressions for diagrams rotated by 90 degrees one has to apply
a Fierz transformation.

\addtolength{\arraycolsep}{-1mm}
\begin{table}[htb]
\begin{displaymath}
\begin{array}{||c||c|c||c|c||c|c||c|c||}
\hline
\hline
ij &
	\multicolumn{2}{c||}{++} &
	\multicolumn{2}{c||}{+-} &
	\multicolumn{2}{c||}{-+} &
	\multicolumn{2}{c||}{--} \\
\hline
\ & \ & \ & \ & \ & \ & \ & \ & \ \\[-4mm]
\ell &
	{\cal C}_\ell^{ij} & {\tilde {\cal C}}_\ell^{ij} &
	{\cal C}_\ell^{ij} & {\tilde {\cal C}}_\ell^{ij} &
	{\cal C}_\ell^{ij} & {\tilde {\cal C}}_\ell^{ij} &
	{\cal C}_\ell^{ij} & {\tilde {\cal C}}_\ell^{ij} \\
\hline
\hline
\ & \ & \ & \ & \ & \ & \ & \ & \ \\[-4mm]
{\rm D_0} &
	1 & N+2 &
	1 & -N &
	1 & -N &
	1 & N-2 \\
\hline
\ & \ & \ & \ & \ & \ & \ & \ & \ \\[-3mm]
{\rm D_1} &
	\frac{2N^2+2N-1}{2N} & - \frac{N+2}{2N} &
	- \frac{1}{2N} & \frac{1}{2} &
	- \frac{1}{2N} & \frac{1}{2} &
	\frac{2N^2-2N-1}{2N} & \frac{-N+2}{2N} \\[2mm]
{\rm D_2} &
	- \frac{1}{2N} & \frac{N^3+2N^2-2}{2N} &
	- \frac{1}{2N} & - \frac{N^2-2}{2} &
	- \frac{1}{2N} & - \frac{N^2-2}{2} &
	- \frac{1}{2N} & \frac{N^3-2N^2+2}{2N} \\[2mm]
{\rm D_3} &
	\frac{N^2+2N-1}{2N} & - \frac{1}{N} &
	- \frac{N^2+1}{2N} & 1 &
	- \frac{N^2+1}{2N} & 1 &
	\frac{N^2-2N-1}{2N} & \frac{1}{N} \\[2mm]
{\rm D_4} &
	- \frac{1}{2N} & \frac{N^3+2N^2-2}{2N} &
	- \frac{1}{2N} & - \frac{N^2-2}{2} &
	- \frac{1}{2N} & - \frac{N^2-2}{2} &
	- \frac{1}{2N} & \frac{N^3-2N^2+2}{2N} \\[2mm]
{\rm D_5} &
	\frac{N^2-1}{2N} & \frac{N^3+2N^2-N-2}{2N} &
	\frac{N^2-1}{2N} & - \frac{N^2-1}{2} &
	\frac{N^2-1}{2N} & - \frac{N^2-1}{2} &
	\frac{N^2-1}{2N} & \frac{N^3-2N^2-N+2}{2N} \\[2mm]
{\rm D_6} &
	\frac{N^2+N-1}{2N} & \frac{N^2-2}{2N} &
	\frac{N^2+N-1}{2N} & - \frac{N}{2} &
	\frac{N^2-N-1}{2N} & \frac{N}{2} &
	\frac{N^2-N-1}{2N} & - \frac{N^2-2}{2N} \\[2mm]
{\rm D_7} &
	\frac{N-1}{2N} & \frac{N^2+N-2}{2N} &
	\frac{N-1}{2N} & - \frac{N-1}{2} &
	- \frac{N+1}{2N} & \frac{N+1}{2} &
	- \frac{N+1}{2N} & - \frac{N^2-N-2}{2N} \\[2mm]
\hline
\hline
\end{array}
\end{displaymath}
\caption[]{Color factors of the diagrams ${\rm D}_\ell^{ij},
	\; \ell=0,\ldots,7$}
\label{ColorTable}
\end{table}
\addtolength{\arraycolsep}{1mm}

%
%

\end{document}